\documentclass[aps,prb,twocolumn,groupedaddress,superscriptaddress,floatfix,longbibliography]{revtex4-2}
\usepackage{epsfig}
\usepackage{multirow}
\usepackage{amsmath,amssymb,mathtools}
\usepackage{color}
\usepackage{graphicx}% Include figure files
\usepackage{dcolumn}% Align table columns on decimal point
\usepackage{bm}% bold math
\usepackage{subfigure}
\usepackage[utf8]{inputenc}
\usepackage[T1]{fontenc}
\usepackage{tikz}
\usepackage{bigints}
\usepackage{orcidlink}
\usepackage{url}
\usepackage{comment}
\usepackage{natbib}

\begin{document}

\title{Driving a Josephson Traveling Wave Parametric Amplifier into chaos: effects of a non--sinusoidal current--phase relation}

\author{Claudio Guarcello\,\orcidlink{0000-0002-3683-2509}}
\email{cguarcello@unisa.it}
\affiliation{Dipartimento di Fisica ``E.R. Caianiello'', Universit\`a di Salerno, Via Giovanni Paolo II, 132, I-84084 Fisciano (SA), Italy}
\affiliation{INFN, Sezione di Napoli Gruppo Collegato di Salerno, Complesso Universitario di Monte S. Angelo, I-80126 Napoli, Italy}
\author{Carlo Barone\,\orcidlink{0000-0002-6556-7556}}
\email{cbarone@unisa.it }
\affiliation{Dipartimento di Fisica ``E.R. Caianiello'', Universit\`a di Salerno, Via Giovanni Paolo II, 132, I-84084 Fisciano (SA), Italy}
\affiliation{INFN, Sezione di Napoli Gruppo Collegato di Salerno, Complesso Universitario di Monte S. Angelo, I-80126 Napoli, Italy}
\author{Giovanni Carapella\,\orcidlink{0000-0002-0095-1434}}
\email{gcarapella@unisa.it }
\affiliation{Dipartimento di Fisica ``E.R. Caianiello'', Universit\`a di Salerno, Via Giovanni Paolo II, 132, I-84084 Fisciano (SA), Italy}
\affiliation{INFN, Sezione di Napoli Gruppo Collegato di Salerno, Complesso Universitario di Monte S. Angelo, I-80126 Napoli, Italy}
\author{Veronica Granata\,\orcidlink{0000-0003-2246-6963}}
\email{vgranata@unisa.it }
\affiliation{Dipartimento di Fisica ``E.R. Caianiello'', Universit\`a di Salerno, Via Giovanni Paolo II, 132, I-84084 Fisciano (SA), Italy}
\affiliation{INFN, Sezione di Napoli Gruppo Collegato di Salerno, Complesso Universitario di Monte S. Angelo, I-80126 Napoli, Italy}
\author{Giovanni Filatrella\,\orcidlink{0000-0003-3546-8618}}
\email{filatr@unisannio.it}
\affiliation{Science and Technology Department, University of Sannio, Benevento, Italy}%
\author{Andrea Giachero\,\orcidlink{0000-0003-0493-695X}}
\email{andrea.giachero@unimib.it }
\affiliation{Department of Physics, University of Milano Bicocca, Piazza
della Scienza, I-20126 Milano, Italy}
\affiliation{INFN—Milano Bicocca, Piazza della Scienza, I-20126 Milano, Italy}
\affiliation{Bicocca Quantum Technologies (BiQuTe) Centre, Piazza della Scienza, I-20126 Milano, Italy}
\author{Sergio Pagano\,\orcidlink{0000-0001-6894-791X}}
\email{spagano@unisa.it }
\affiliation{Dipartimento di Fisica ``E.R. Caianiello'', Universit\`a di Salerno, Via Giovanni Paolo II, 132, I-84084 Fisciano (SA), Italy}
\affiliation{INFN, Sezione di Napoli Gruppo Collegato di Salerno, Complesso Universitario di Monte S. Angelo, I-80126 Napoli, Italy}

\date{\today}

\begin{abstract}
We tackled the numerical analysis of the dynamic response of a Josephson Traveling Wave Parametric Amplifier (JTWPA) by varying the driving parameters, with a focus on the pathways leading to chaotic behavior. By tuning the working conditions, we explore the broad spectrum of dynamical regimes accessible to JTWPAs, delineating the conditions under which transition to chaos occurs. Furthermore, we extend our investigation to junctions characterized by a non--sinusoidal current phase relation (CPR) and explore the consequences on the amplifier's performance.
Through the study of gain characteristics, Poincar\'e sections, and Fourier spectra, we provide an in-depth understanding of how CPR nonlinearity and nonsinusoidality influence the operational effectiveness and stability of JTWPAs. This investigation offers insights into optimizing the device for enhanced performance and robustness against chaotic disruptions, in order to establish a framework for predicting and controlling JTWPA behavior in practical applications. In fact, we identify the regions in the parameter space where the input signal is maximally amplified without excessive noise or undesired harmonics. This effort paves the way for the development of devices with tailored dynamic responses and for advancements in quantum computing and precision measurement technologies, where stability and high fidelity are of paramount importance. 
\end{abstract}

\maketitle

\section{Introduction}
\label{Sec1}

\begin{figure*}[t!!]
\centering
\includegraphics[width=1.75\columnwidth]{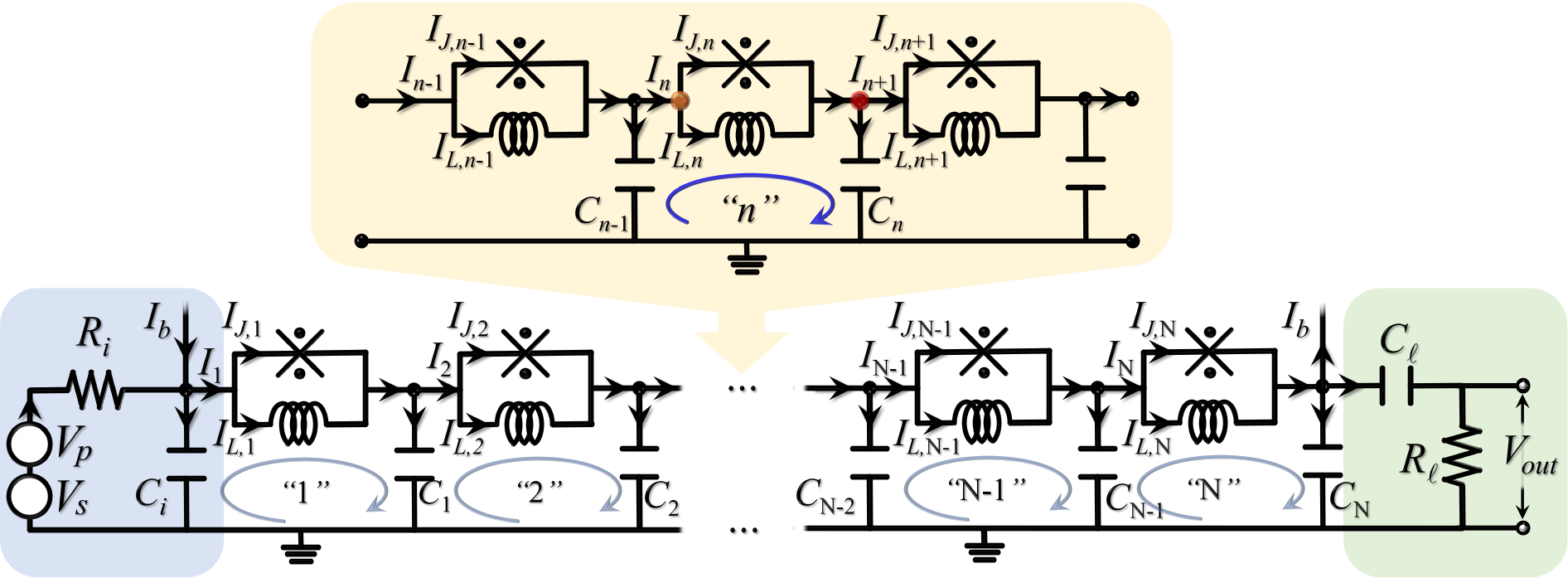}
\caption{Outline of the JTWPA electrical circuit: the bottom part contains the input, cyan shaded, and output, green shaded, cells, encompassing the pump and signal tones, $V_{\text{pump}}$ and $V_{\text{sign}}$, respectively, the input and load resistances and capacitances, denoted as $R_i$, $C_i$, $R_{\ell}$, and $C_{\ell}$. The output voltage is measured across the load resistor. The top part highlights, within a shaded yellow box, an in--depth looking at a cell of the series of $N = 990$ elements of the device.}
\label{Fig01}
\end{figure*}

Amplifying weak microwave signals is crucial across numerous applications, including readouts for superconducting qubits, quantum devices, and radio astronomy~\cite{Bartram2023}. Leading low--noise amplifiers in the microwave domain approach the quantum limited noise performance by operating at mK temperatures and parametric pumping of nonlinear circuits is achieved with Josephson junctions (JJs) or high--kinetic--inductance components~\cite{Aumentado2020}.

While nonlinear resonator--based parametric amplifiers typically have bandwidths in the tens of MHz range, constrained by gain and resonator linewidth, \emph{traveling-wave parametric amplifiers} (TWPAs) offer much broader bandwidths, reaching several GHz~\cite{Cullen1960,Sorenssen1962,Esposito2021}. TWPAs' increased bandwidth enables extensive frequency multiplexing, particularly beneficial for qubit readouts and single-photon detectors~\cite{Pagano2022}. Efficiently using hardware resources through multiplexed approaches is crucial for large quantum devices.
Challenges in TWPAs include material losses and the generation of signal sidebands, identified as primary sources of excess noise beyond the quantum limit. In addition to minimizing noise at the signal frequency, broadband amplifiers must avoid generating spurious tones due to intermodulation of inputs, especially in frequency-multiplexed scenarios.

Presently, superconducting TWPAs employ two sources of nonlinearity. Firstly, through the inherent distributed nonlinear kinetic inductance found in superconductors~\cite{Bockstiegel2014,Vissers2016,Chaudhuri2017,Ranzani2018,Malnou2021}. Alternatively, one can incorporate JJs into the transmission line, thus introducing a nonlinear inductance element, obtaining the so-called Josephson TWPAs (JTWPAs)~\cite{Sweeny1985,Bell2015,Zorin2016,Fasolo2019,Esposito2021,Esposito2022}. 
In this study, we focus on JTWPAs, using the specific device details from the DARTWARS INFN collaboration framework~\cite{Giachero2022,Pagano2022,Granata2023,Borghesi2023,Rettaroli2023,Fasolo2024,Ahrens2024,Faverzani2024} as our baseline. In particular, we use the specifications given in Ref.~\cite{Pagano2022} to theoretically explore the response of this device to the simultaneous application of pump and signal excitations. Our aim is to scan the parameter space to optimize experimental conditions that maximize the gain, calculated as the amplification of the output signal at the frequency of the input signal tone. To do so, it is essential to avoid the noise contribution from the chaotic states. In fact, the pump tone and the bias current flowing through the system can exceed the threshold parameter values above which the system enters a regime of chaotic response, which is practically useless for the amplification applications we are interested in. In this way, we can define the system parameter ranges that maximise the gain provided by JTWPA without incurring detrimental effects. Fourier transform (FT) and Poincar\'e section (PS) analysis are our primary investigative tools to unravel the transitions to chaos. This choice is very effective in capturing dynamic spectral properties and the geometrical structure of phase space, respectively. While alternative methods, like the study of Lyapunov exponents, can offer insights into JJ stability and chaotic behavior~\cite{Shukrinov2012,Botha2013,Shukrinov2014,Sokolovic2017}, FT and PSs provide a quite intuitive understanding of the complex interactions and transitions in JTWPAs.

The paper is organized as follow. In Sec.~\ref{Sec2}, we introduce the theoretical approach used to model the behavior of the JTWPA, and the values we set for the system parameters. In Sec.~\ref{Sec3}, we present the analysis as a function of the pump intensity, signal frequency, and bias current. In Sec.~\ref{Sec4}, we show how the transparency of the junction, i.e., the skewness of the current-phase relation (CPR), impacts on the device performances. In Sec.~\ref{Sec5}, conclusions are drawn.

\section{Model} 
\label{Sec2}

We deal with the behavior of the JTWPA sketched in Fig.~\ref{Fig01}~\cite{GuarcelloJTWPA2023,GuarcelloJTWPA2024}: the parameters for the system, discussed herein, are derived in Ref.~\onlinecite{Pagano2022}. The yellow-shaded part of Fig.~\ref{Fig01} focus on the $n-$th cell of the 990 cells of the JTWPA, each of which incorporates an rf--SQUID, made by a JJ in parallel with a geometric inductance $L_{g,n} = 120\, \text{pH}$, and a capacitance $C_{g,n} = 24\, \text{fF}$. The input cell (highlighted in blue in Fig.~\ref{Fig01}) includes a pump and signal voltage source, denoted as $V_{\text{pump}}$ and $V_{\text{sign}}$, along with an input resistance and capacitance $R_i = 50\, \Omega$ and $C_i = 24\, \text{fF}$, respectively. The output cell (highlighted in green in Fig.~\ref{Fig01}) is supplied with a load resistance and capacitance, $R_{\ell} = 50\, \Omega$ and $C_{\ell} = 1\, \text{nF}$, respectively.

Details of the numerical model that we employed are given in App.~\ref{AppA}. Within the framework of the Resistively and Capacitively Shunted Junction (RCSJ) model~\cite{Barone1982}, the Josephson current through the $n$-th junction can be written as
\begin{equation}
I_{J,n} = C_J \frac{\hbar}{2e} \frac{d^2 \varphi_n}{dt^2} + \frac{1}{R_J} \frac{\hbar}{2e} \frac{d\varphi_n}{dt} + I_c \sin(\varphi_n).
\end{equation}
Here, we assume that each JJ has a critical current $I_c = 2\, \mu \text{A}$, a quasiparticle resistance $R_J = 20\, k\Omega$, and a capacitance $C_J = 200\, \text{fF}$, thus yielding a plasma oscillation frequency $\nu_p = 27.7\, \text{GHz}$. The phase dependence of the Josephson current is firstly taken sinusoidal, i.e., $I(\varphi) = I_c \sin(\varphi)$; then, in Sec.~\ref{Sec4}, this assumption will be relaxed where a transparency--dependent non--sinusoidal CPR is considered. From the second Josephson equation, $V = \frac{\hbar}{2e} \frac{d\varphi}{dt}$, one obtains a current-voltage profile characteristic for a nonlinear inductor, $V = L_{J_0}(\varphi) dI/dt$,
where
\begin{equation}
L_{J_0}(\varphi) = \frac{\hbar}{2e} \frac{1}{I_c \cos(\varphi)}
\end{equation}
denotes the intrinsic Josephson inductance. Considering a non-zero DC bias current, $I_{\text{bias}}$, the effective Josephson inductance adjusts to $L_J(\varphi) = L_{J_0}(\varphi)\bigg/\sqrt{1 - \left(\frac{I_{\text{bias}}}{I_c}\right)^2}$.
Thus, the bias current serves to regulate the Josephson inductance, enhancing it as $I_{\text{bias}}$ enlarges, until it diverges as it approaches $I_c$.
The role of the current--dependent Josephson inductance is crucial in the process of parametric mixing and amplification, in which a high amplitude pump tone at frequency $\nu_{\text{pump}}$ and a weaker signal at frequency $\nu_{\text{sign}}$ are employed. The large amplitude of the pump tone regulates the nonlinear characteristics of the Josephson inductance, promoting an interaction that can lead to the amplification of the signal and, at the same time, the generation of an \emph{idle tone} at frequency $\nu_{\text{idle}}$. The nature of this interaction is significantly influenced by the way the non-linearity of the inductance varies, leading to two main modes of operation.
In the four--wave mixing (4WM) regime, the relation between the operating frequencies is given by $2 \nu_{\text{pump}} = \nu_{\text{sign}} + \nu_{\text{idle}}$. On the other hand, the three--wave mixing (3WM) mode is described by the relation $\nu_{\text{pump}} = \nu_{\text{sign}} + \nu_{\text{idle}}$. The transition to the 3WM regime is often due to an odd current dependence in the inductance nonlinearity, typically achieved by the application of a DC bias and/or a magnetic field.
This distinction emphasizes the importance of the study of the current-dependent Josephson inductance nonlinearity in determining the operating mode and the efficiency of the parametric amplification mechanism.

Theoretical approaches generally linearise the Josephson inductance~\cite{Yaakobi2013,O'Brien2014}, towards a simplified scenario with lower-order approximations. These analytical treatments, based on the so-called \emph{coupled mode equations} and their subsequent modifications~\cite{Chaudhuri2015,Zorin2016,Dixon2020,Kern2023}, give valuable insight into the physics of such devices, but fail to accurately describe all possible scenarios. Thus, recently, other numerical methods have been efficiently pursued, often using open--source tools~\cite{Dixon2020,Sweetnam2022,Gaydamachenko2022,OPeatain2023,Levochkina2024,Krause2024,Elkin2024}. Instead, our approach tackles the challenge comprehensively by solving a system of coupled differential equations, one for each cell constituting the JTWPA, along with appropriate boundary conditions. The choice of boundary conditions is quite crucial when studying JJ chains; for example, it has been shown how load-matched boundary conditions can significantly change the dynamics and also suppress chaos~\cite{Pankratov2017,Pankratov2024}. The system's dynamics is derived by integrating these differential equations, employing an implicit finite difference method grounded in a tridiagonal algorithm -- a common choice for investigating Josephson transmission lines numerically (see App.~\ref{AppA} for a comprehensive description) -- and examining the spectral components of the output signal through the Mathematica software. The parameters for the time step and integration period, normalized to the Josephson plasma frequency, are chosen as $\Delta t = 0.01$ and $t_{\text{max}} = 20000$, respectively; these values translate to $\Delta t = 0.06$ ps and $t_{\text{max}} = 120$ ns in physical units. This configuration ensures a wide enough duration for the integration process, effectively mitigating transient effects and maintaining the numerical integrator's stability, thereby facilitating a robust analysis of steady-state behaviors. 

The different operating method, in addition of giving complete control over the effect of individual parameters and their variations, allows to modify the intrinsic characteristics of the JJs. In fact, we will also look at the system's response if, instead of a sinusoidal CPR, we take into account a certain junction transparency, which induces non-sinusoidal components in the Josephson current. 

\begin{figure*}[t!!]
\centering
\includegraphics[width=2\columnwidth]{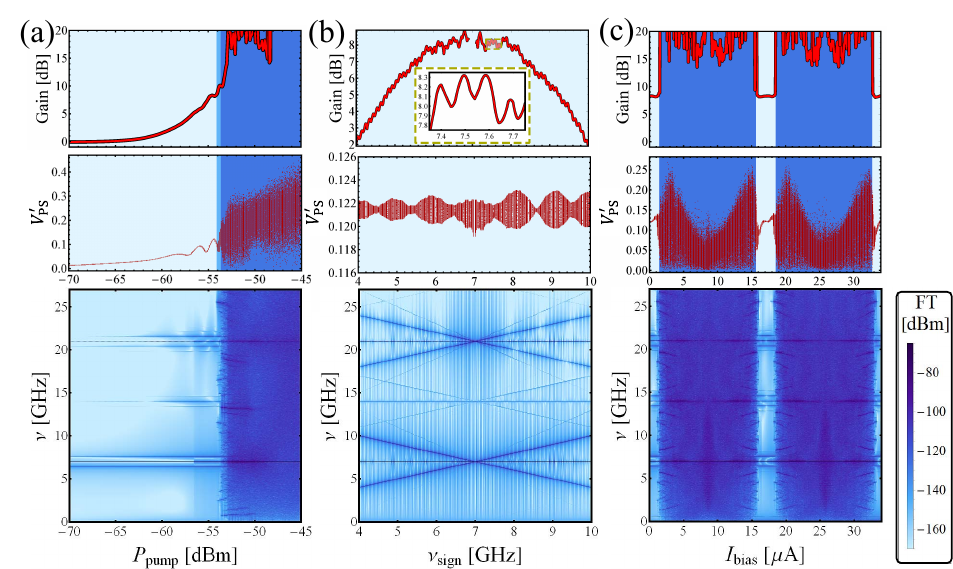}
\caption{Gain (top panel), PS (middle panel), and Fourier spectra of the output voltage (bottom panel) \emph{versus}: (a) pump power level, $P_{\text{pump}}$, at $\nu_{\text{sign}}=6.42\;\text{GHz}$ and $I_{\text{bias}}=0$; (b) the signal frequency, $\nu_{\text{sign}}$, at $P_{\text{pump}}=-55\;\text{dBm}$ and $I_{\text{bias}}=0$; (c) bias current, $I_{\text{bias}}$, at $P_{\text{pump}}=-55\;\text{dBm}$ and $\nu_{\text{sign}}=6.42\;\text{GHz}$. In top and middle panels, different response regimes are highlighted by regions shaded in different colors. In the density plots, the color intensity scale represents the amplitude of the spectral components whose frequency can be read on the bottom-left vertical axis. The other parameters are: $\nu_{\text{pump}}=7\;\text{GHz}$ and $P_{\text{sign}}=-100\;\text{dBm}$.}
\label{Fig02}
\end{figure*}

\section{Gain profiles in the case of a sinusoidal CPR}
\label{Sec3}

Figure~\ref{Fig02} presents the JTWPA dynamics by scanning the pump tone amplitude (a), the frequency of the signal to be amplified (b), and the bias current (c). In particular, we plot the gain (top panel), PS (middle panel), and Fourier spectra of the output voltage (bottom panel) \emph{versus}: (a) pump power level, $P_{\text{pump}}$, at $\nu_{\text{sign}}=6.42\;\text{GHz}$ and $I_{\text{bias}}=0$; (b) signal frequency, $\nu_{\text{sign}}$, at $P_{\text{pump}}=-54.5\;\text{dBm}$ and $I_{\text{bias}}=0$; (c) bias current, $I_{\text{bias}}$, at $P_{\text{pump}}=-55\;\text{dBm}$ and $\nu_{\text{sign}}=6.42\;\text{GHz}$. The signal intensity is set at $P_{\text{sign}} = -100\;\text{dBm}$, equivalent to a voltage amplitude of $V_{\text{sign}} = 0.032$ mV, while the frequencies for pump is maintained at $\nu_{\text{pump}} = 7\;\text{GHz}$. 

The signal gain is defined as $\text{Gain} = 20 \log [V_{\text{out}}(\nu_{\text{sign}})/V_{\text{sign}}]$, while the PSs are constructed by collecting, for each combination of parameters, all values of $V_{\text{out}}'$ at $V_{\text{out}}=0$ (for this reason we label this quantity with $V'_{\text{PS}}$). If we think at a \emph{phase portrait}, $V_{\text{out}}'$ \emph{versus} $V_{\text{out}}$, a narrow distribution of $V'_{\text{PS}}$ values is the result of an elliptical limit cycle, while we expect that the $V'_{\text{PS}}$ distribution of values enlarges in the case of a phase portrait consisting of a closed, tangled skein~\cite{GuarcelloJTWPA2023}.

In Fig.~\ref{Fig02}(a) we explore the impact of the pump intensity, $P_{\text{pump}}$, at $\nu_{\text{sign}} = 6.42\;\text{GHz}$ in the absence of bias current, thus setting the working mode within the 4WM domain, so that an idle tone at $\nu_{\text{idle}} = 7.58\;\text{GHz}$ is expected. Under these conditions, signal amplification is observed throughout the investigated pump intensity range. Distinct amplification regimes are identified, demarcated in Fig.~\ref{Fig02}(a) by different shades of cyan--blue. For $P_{\text{pump}} \lesssim -54\;\text{dBm}$, a moderately increasing gain is noted, with $\text{Gain}\lesssim 8\;\text{dB}$. Within the range $P_{\text{pump}} \in (-54, -53.5)\;\text{dBm}$, the gain shows a sort of jump to $\text{Gain} \sim 10\;\text{dB}$. Beyond this intensity, i.e., for $P_{\text{pump}} \gtrsim -53.5\;\text{dBm}$, the gain assumes quite high values, although further analysis suggests that these conditions do not actually give signal amplification. This also reflects in the behavior of $V'_{\text{PS}}$, see middle panel of Fig.~\ref{Fig02}(a). This plot permits to identify transition regions where the behavior of the system changes markedly, pointed by sudden changes in the density or pattern of points, indicating bifurcations or transitions between different dynamical states. In fact, if we observe how the points are distributed in the $V'_{\text{PS}}$ plot, for $P_{\text{pump}} \lesssim -54\;\text{dBm}$ we note concentrated clusters of points that might indicate stable periodic orbits in the phase portrait; conversely, for higher value of $P_{\text{pump}}$ more spread--out points suggest transitions to chaos, see $P_{\text{pump}} \in (-54, -53.5)\;\text{dBm}$, and the definitive settling of a chaotic behavior, see $P_{\text{pump}} \gtrsim -53.5\;\text{dBm}$. Thus, increasing the pump intensity leads to more complex patterns, suggesting stronger nonlinear interactions within the JTWPA; this might correspond to an enhance of the amplifier's gain, but also a potential increase in instability or noise. Regions within the $V'_{\text{PS}}$ plot where the system appears stable over a range of pump intensities are important for practical application, indicating configurations where the JTWPA can operate efficiently and reliably. 
To better illustrate the effects of changing the pump intensity, we include in the Supplementary Material an animation showing the evolution of the FSs, the phase portraits, and the Josephson phase as $P_{\text{pump}}$ increases. It is evident that the range of values within the Josephson phase oscillates enlarges with $P_{\text{pump}}$, so that $|\varphi|\lesssim1.5$ just before the onset of a chaotic regime; this threshold value is in line with the maximum phase value beyond which the system may have unpredictable behaviour discussed in Ref.~\cite{Kissling2023}.

An in--depth analysis of the output signal's temporal evolution is also warranted, e.g., looking at the Fourier analysis of $V_{\text{out}}(t)$, as depicted in the bottom panel of Fig.~\ref{Fig02}(a). This map shows the presence of different frequency components within the output signal, where the peaks indicate the dominant ones. 
%The evolution of these peaks across pump intensities reveals how the JTWPA's amplification characteristics change with $P_{\text{pump}}$.
At low pump powers, the spectrum is relatively sparse, with only a few harmonics, suggesting linear or weakly nonlinear behavior where the JTWPA slightly amplifies the signal without significantly altering its frequency content.
As $P_{\text{pump}}$ increases, we notice more spectral components, but also the emergence/strengthening of harmonic peaks, indicating nonlinear effects like harmonic generation.
This is a crucial aspect of parametric amplification, where the nonlinearity of the JJs can generate new frequency components not present in the input signal. However, changing $P_{\text{pump}}$ leads to a spread of the frequency spectrum and modifications of the output signal bandwidth. 
A broader spread might suggest an increase in the amplifier's effective bandwidth at higher pump powers, or it could indicate increased noise or nonlinear distortion due to the onset of a chaotic response.
Indeed, at certain pump power levels, i.e., for $P_{\text{pump}} \gtrsim -53.5\;\text{dBm}$, there is a sudden proliferation of new spectral lines and a noticeable broadening of the existing ones. This is typically indicative of a period--doubling cascade, i.e., a common route to chaos~\cite{Levinsen1982,Olsen1985,Shukrinov2014}. 
%The densest areas, with a quite noisy spectrum, indicate fully developed chaotic behavior, where the amplifier generates a wide range of frequencies due to highly nonlinear and aperiodic dynamics.
Thus, clear, distinct peaks suggest a more stable amplification process, while a very dense, broad distribution of frequency components might indicate a higher noise level, i.e., for $P_{\text{pump}} \gtrsim -53.5\;\text{dBm}$ the system's dynamics undergo a stark transition to chaos. 
%By identifying regions where the desired signal components are maximally amplified without excessive noise or undesired harmonics, one can pinpoint optimal operational conditions for the JTWPA. These are pump intensity values that offer the best balance between amplification and signal fidelity. 

In Fig.~\ref{Fig02}(b) we inspect the effects produced by a variation of the signal frequency. 
In the top panel, the amplifier's gain \emph{versus} signal frequency characteristic is illustrated under a ``clean'' gain regime, but in working conditions close to the onset of a chaotic response, so as to maximise the gain achieved, i.e., we impose $P_{\text{pump}} = -54.5\;\text{dBm}$. 
We achieve $\text{Gain}\sim 8\;\text{dB}$ in the $\nu_{\text{sign}}\in[6-8]\;\text{GHz}$ frequency bandwidth. Along this gain profile, periodic variations, usually referred to as ``ripples'', are prominently observed (refer to the detailed view in the inset); each ripple is associated in this case with small gain variations of $\sim0.2\;\text{dB}$. Common to all systems with impedance mismatches, a ripple forms along the length of the waveform in the array due to complex interference between the forward and backward propagating waves encountering multiple reflections~\cite{Peatain2023,Zheng2024}. This ripple has nodes along the length of the array for every half-wavelength period of the signal tone, while the maximum of the signal gain between nodes depends on the particular wave mixing dynamics in the transmission line~\cite{Kern2023}. This emerging ripple pattern depends on parameter values and can influence both the gain and the bandwidth of the device~\cite{GuarcelloJTWPA2024}, although these effects can be mitigate by implementing dispersion engineering strategies~\cite{O'Brien2014,White2015,Macklin20215,Winkel2020,Planat2020,Gaydamachenko2022,Roudsari2023,Qiu2023}.
%However, these quantities can be significantly improved through dispersion engineering, e.g., by a \emph{resonant phase matching} strategy~\cite{O'Brien2014,White2015,Macklin20215}. 
%The latter includes periodically embedded LC resonators capacitively coupled to the transmission line: the resonators adapt the dispersion relation so that the total phase mismatch along the transmission line remains small over a wide bandwidth. 
%However, also other dispersion engineering strategies can be taken into account for maximizing the gain achievable and the robustness of these devices~\cite{Winkel2020,Planat2020,Gaydamachenko2022,Roudsari2023,Qiu2023}. 
However, in this work we are mainly interested on finding threshold regimes, which allow reliable operation without the onset of any chaotic response, and on the analysis of the anhamonic effects produced by a non-sinusoidal CPR, so we consider a JTWPA without any dispersion--engineering strategy.

The PS plot in middle panel of Fig.~\ref{Fig02}(b) shows, this time, quite dense clusters of points, corresponding to stable periodic orbits in a phase portrait; in other words, a change of frequency alone is not able to induce a transition to a chaotic regime. This is also evident when looking at the Fourier analysis of the output signal presented in the bottom panel of Fig.~\ref{Fig02} for $\nu_{\text{sign}}\in[4-10]\;\text{GHz}$. This map prominently features peaks associated with the pump, signal, and idler frequencies across the entire scanned range, including their first and second harmonics. This observation underscores the amplifier's extensive bandwidth. Moreover, additional spectral lines are visible, corresponding to various frequency combinations arising during the operation. These ``extraneous'' tones detract from the power of the signal to be amplified, potentially leading to system instability. Notably, vertical striations in this density plot are linked to the presence of gain ripples discussed above.

\begin{figure*}[t!!]
\centering
\includegraphics[width=1.5\columnwidth]{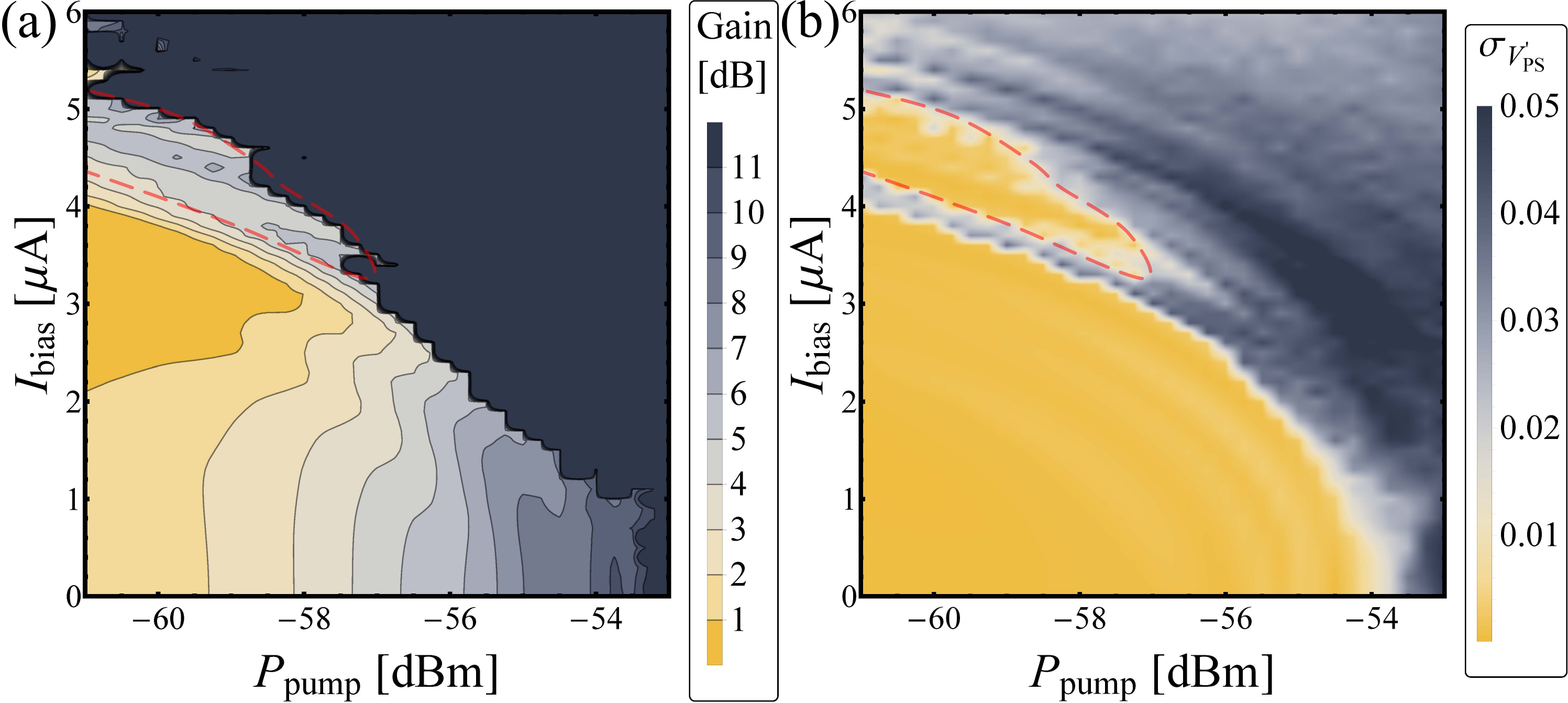}
\caption{(a) Gain and (b) $V'_{\text{PS}}$--standard deviation, $\sigma_{V'_{\text{PS}}}$, as a function of $I_{\text{bias}}\in[0,6]\;\mu\text{A}$ and $P_{\text{pump}}\in[-61,-53]\;\text{dBm}$. The other parameters are: $\nu_{\text{sign}}=6.42\;\text{GHz}$, $\nu_{\text{pump}}=7\;\text{GHz}$, and $P_{\text{sign}}=-100\;\text{dBm}$.}
\label{Fig03}
\end{figure*}

Figure~\ref{Fig02}(c) illustrates the effects produced by a bias current flowing through the JTWPA. It modifies the Josephson inductance nonlinearity, making the system more receptive to 3WM processes. This can enhance the device's capability to mediate energy transfer between different frequency components effectively. Indeed, the presence of both 3WM and 4WM modes allows for more versatile frequency conversion capabilities: 3WM can complement 4WM processes by facilitating additional pathways for energy transfer and signal amplification. 
Looking at the gain (top panel) and the PS (middle panel) profiles \emph{versus} $I_{\text{bias}}$, we recognize regions of bias current values wherein the response of the system is stable, with $\text{Gain}\sim 8\;\text{dB}$, but also clear transitions to a chaotic regime. Nevertheless, the system seems to display the same behavior every $\sim 17\,\mu A$. We observe that the phase matching across the rf--SQUID arms leads to the condition
\begin{equation}
 2\pi\frac{L_g I_g}{\Phi_0} = \varphi.
\end{equation}
When $\varphi = 2\pi n$, with $n=0,\pm1,\pm2\dots$, the current through the inductance is $I_g = n \Phi_0/L_g$.
Put another way, in the case of $L_{g,n} = 120\, \text{pH}$, a bias current equal to $I^{2\pi}_{\text{bias}}=n\Phi_0/L_g \approx n\times17.1\,\mu A$ induces $2\pi n$-phase rotations, thus restoring the initial working conditions. This mechanism explains the observed periodicity upon varying the bias current. To switch between two stable configurations with a different $n$, the system has to undergo through a $2\pi$--phase rotation and an unstable condition, corresponding to the discussed chaotic response. This beahvior emerges clearly in the animation attached to Supplementary Materials.

The periodicity described so far is quite evident also from the FT map in the bottom panel of Fig.~\ref{Fig02}(c); the alternation of stable/unstable zones is clear. At lower bias currents, the output spectrum is more ordered, with distinct spectral lines that show a predictable response to changes in $I_{\text{bias}}$, i.e., a stable regime where the Josephson inductance is being modulated in a controlled way.
As the bias current moves away from this stable region, the spectra become increasingly complex and disordered. There are regions with a dense distribution of spectral lines, which could be evidence of a quasiperiodic route to chaos--where the introduction of new frequency components leads to a breakdown of periodicity.
At the furthest extents, the $I_{\text{bias}}$-dependent FT map shows a very dense spectrum, suggesting that the dynamics of the JTWPA have become aperiodic and chaotic. 
%This regime is characterized by a large number of frequencies with significant power, indicating a breakdown of the periodic motion that instead dominated the stable region.
Figure~\ref{Fig02}(c) shows also that there is an entire window of $I_{\text{bias}}$ values, i.e., $\Delta I_{\text{bias}}=I^{2\pi}_{\text{bias}} \pm \delta I_{\text{bias}}$ with $\delta I_{\text{bias}}\simeq 1.5\;\mu\text{A}$, within which the system shows a stable response. A close look at this map also reveals that at low frequencies the characteristic peak of the 3WM mode at $\nu_{\text{idle}} = 0.58\;\text{GHz}$ tends to emerge.

It is reasonable to expect that the $I_{\text{bias}}$ range of values that ensures the stability of the system may depend on the intensity of the pump tone, i.e., we can inspect how $\delta I_{\text{bias}}$ varies with $P_{\text{pump}}$. Since the behavior at $n\neq0$ is a replica of what happens around $I_{\text{bias}}=0$, we focus on the system's response to small bias currents. Thus, in Fig.~\ref{Fig03} we show (a) the gain and (b) the $V'_{\text{PS}}$--standard deviation, $\sigma_{V'_{\text{PS}}}$, as a function of $I_{\text{bias}}\in[0,6]\;\mu\text{A}$ and $P_{\text{pump}}\in[-61,-53]\;\text{dBm}$. Specifically, in Fig.~\ref{Fig03}(a) we map $\text{Gain}(P_{\text{pump}},I_{\text{bias}})$: in this graph, it is possible to identify a threshold, above which a dark grey area marks parameter combinations that give quite high gains, indicative of a chaotic regime. Below this threshold, we see that: 
\emph{i}) at fixed $I_{\text{bias}}$, increasing $P_{\text{pump}}$ increases the gain;
\emph{ii}) at fixed $P_{\text{pump}}$, increasing $I_{\text{bias}}$ induces non-monotonic gain trends. For instance, at $P_{\text{pump}}=-59\;\text{dBm}$ the gain goes from $\sim3$, then drops to $0$, then rises again to $\sim6$ just before entering a chaotic regime. Alongside the gain map, we also construct a density plot showing how the clusters of PS points distribute; in particular, we calculate for each parameter combination the standard deviation of the $V'_{\text{PS}}$ distribution, $\sigma_{V'_{\text{PS}}}$, see Fig.~\ref{Fig03}(b). In this way, a stable response will correspond to a very small value of $\sigma_{V'_{\text{PS}}}$ (i.e., $\sigma_{V'_{\text{PS}}}\ll0.1$, see the yellow area), while a transition to chaos is marked by a sudden increase in $\sigma_{V'_{\text{PS}}}$ (i.e, $\sigma_{V'_{\text{PS}}}\gtrsim0.1$, see the gray shaded area). Moreover, the $\sigma_{V'_{\text{PS}}}(P_{\text{pump}},I_{\text{bias}})$ map allow to recognize a peculiarity emerging at low $P_{\text{pump}}$'s: in fact, as the current increases, the system first enters a chaotic phase (demarcated by a thin gray belt), and then it lands in a region of parameter space with a stable response (and high gains), marked by a red dashed line containing a yellow region at high currents, i.e., at $I_{\text{bias}}\in(4,5)\;\mu\text{A}$. 

\section{The 3WM and 4WM nonlinear coefficients}
\begin{figure*}[t!!]
\centering
\includegraphics[width=0.8\textwidth]{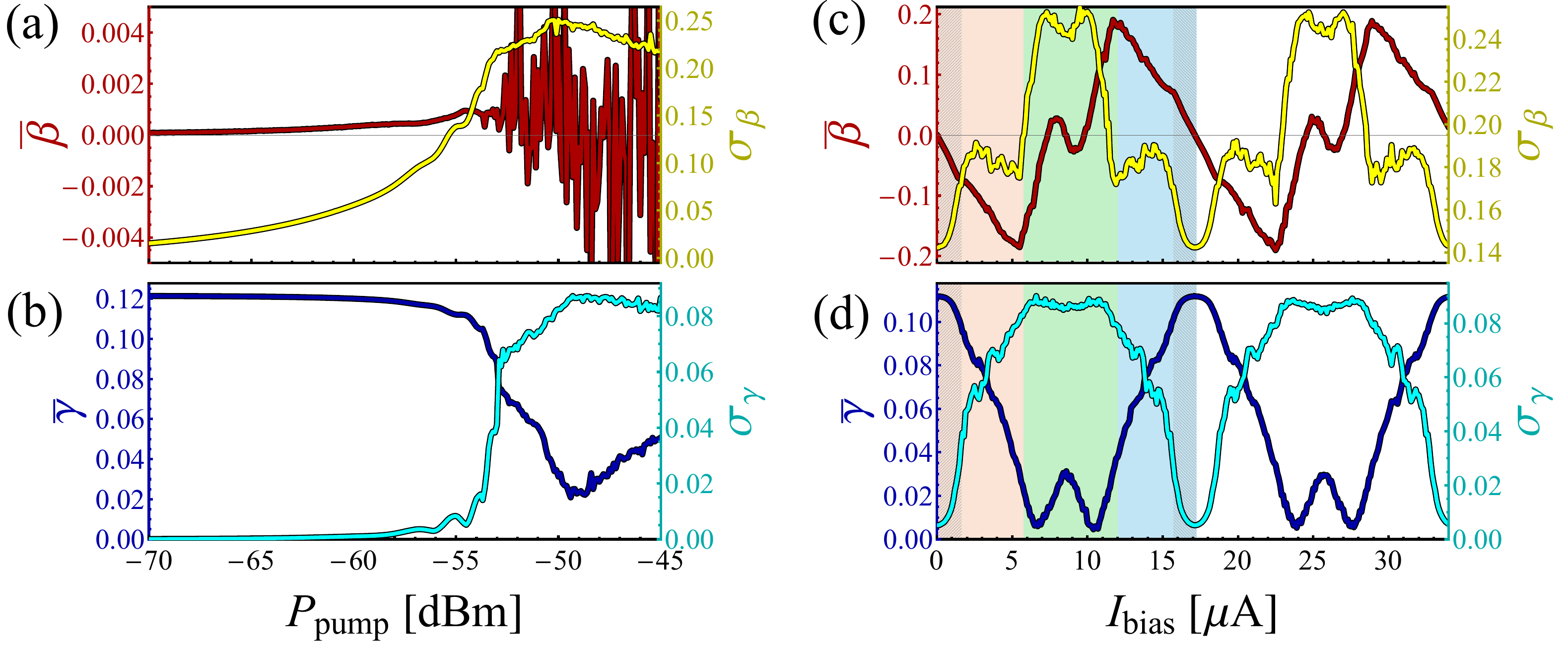}
\caption{Average value (left vertical axes) and standard deviation (right vertical axes) of $\beta(t)$ and $\gamma(t)$ by varying $P_{\text{pump}}$ at $I_{\text{bias}} = 0$ (see left panels) and varying $I_{\text{bias}}$ at $P_{\text{pump}} = -55$ dBm (see right panels). The other parameters are: $\nu_{\text{sign}}= 6.42$ GHz, $\nu_{\text{pump}}= 7$ GHz, and $P_{\text{sign}} = -100$ dBm.}
\label{Fig07}
\end{figure*}

In this section we investigate the behavior of the 3WM and 4WM nonlinear coefficients~\cite{Zorin2016}, respectively given by
\begin{equation}
\beta = \frac{\beta_L}{2} \sin(\varphi_{\text{dc}})
\qquad\text{and}\qquad
\gamma = \frac{\beta_L}{6} \cos(\varphi_{\text{dc}}),
\end{equation}
where $\beta_L = 2\pi L I_c/\Phi_0$ is the screening parameter of the rf-SQUID (in our case, $\beta_L \simeq0.74$) and $\varphi_{\text{dc}} = 2\pi \Phi_{\text{dc}}/\Phi_0$, with $\Phi_{\text{dc}}$ being the magnetic flux inside the loop. The parameter $\beta$ is associated with the quadratic nonlinearity and plays a role in controlling the JTWPA signal gain. An optimal value of $\beta$ allows maximizing the gain while minimizing the influence of cubic nonlinearity~\cite{Zorin2016}. This is particularly beneficial in the 3WM regime, where significant quadratic nonlinearity is necessary to achieve exponential gain. The parameter $\gamma$ represents the cubic (Kerr-like) nonlinearity of the system and affects the phase mismatch and both self-phase and cross-phase modulations.

In Zorin's work~\cite{Zorin2016}, these coefficients were computed considering an external magnetic flux. Since we have no external magnetic drive, we focus on the magnetic flux induced by the circulating current, $I_{\text{circ}}$, in the rf-SQUID loop, i.e., $\Phi_{\text{dc}} = L_{g} I_{\text{circ}}$. This current is determined by the balance between those flowing in the upper and lower branches of the loop, i.e., $I_{\text{circ}} = I_{L,n}-I_{J,n}$ in the $n$-th rf-SQUID of the chain, see App.~\ref{AppA}.

To study the dependence of the 3WM and 4WM coefficients on the pump power $P_{\text{pump}}$ and bias current $I_{\text{bias}}$, we have computed the time evolution of $\varphi_{\text{dc}}(t)$, from which we obtain both $\beta(t)$ and $\gamma(t)$, looking at the last rf-SQUID of the JTWPA, i.e., $n\equiv N$. The behavior of these quantities is shown in the animations included in the Supplementary Materials. We observe that $\beta(t)$ oscillates around a mean value of zero, which agrees with Zorin's results in the absence of an external magnetic field~\cite{Zorin2016}. In contrast, $\gamma(t)$ has initially a mean value around $0.12$, but, as $P_{\text{pump}}$ increases and the system enters into a chaotic regime, it also fluctuates significantly and eventually averages to nearly zero. Given the strongly oscillating behaviour of these quantities, it is more convenient to extract only the mean value and the standard deviation of $\beta(t)$ and $\gamma(t)$ from each time evolution. In this way, it is possible to plot both $\bar{\beta}$, $\sigma_\beta$, $\bar{\gamma}$, and $\sigma_\gamma$ as $P_{\text{pump}}$ and $I_{\text{bias}}$ vary, see Fig.~\ref{Fig07}. As $P_{\text{pump}}$ increases, $\bar{\beta}$ (red curve) remains close to zero, but $\sigma_\beta$ (yellow curve) tends to increase, due to the increasing fluctuations of $\beta(t)$. Once the system enters a chaotic regime, also $ \bar{\beta}$ begins to significantly fluctuate. Instead, $\bar{\gamma}$ (blue curve) remains quite close to the value $\beta_L/6$ as long as the system remains in a non-chaotic condition, after which its value tends to decrease significantly, while at the same time  $\sigma_\gamma$ (cyan curve) increases.

In Fig.~\ref{Fig07}(c-d), we scan $I_{\text{bias}}$ at a fixed $P_{\text{pump}} = -55$ dBm. One still observes the periodicity noticed in Fig.~\ref{Fig02}, but this time we can also appreciate some trends that emerge at those bias currents that set a fully chaotic response. For example, let us focus on the range $I_{\text{bias}}\in[0,17]$, corrisponding to $\varphi\in[0-2\pi]$, which can be further divided into three intervals highlighted by different colours, see Fig.~\ref{Fig07}(c-d). These panels include also two small strips with gray oblique shading, where the system exhibits a stable dynamics, in contrast with the surrounding areas dominated by chaotic behavior.
Within the highlighted bias current range, $ \bar{\beta}$ switches from positive to negative values: this indicates alternating 3WM contributions as the bias current is varied. Specifically, within the orange--shaded area, $  \left| \bar{\beta} \right|$ increases significantly, while its standard deviation reaches a sort of plateau; insted, $ \bar{\gamma}$ decreases to almost zero, while its standard deviation increases. The average values of $\beta$ and $\gamma$ seem to follow similar patterns in the orange and blue regions, but with opposite trends. This is due to the $2\pi$--phase twist occurring around $I_{\text{bias}} \simeq 17 \, \mu\text{A}$. After this twist, the system enters a phase regime where the nonlinear coefficients $\beta$ and $\gamma$ exhibit similar patterns to those just before the twist, but with opposite signs, resulting in the mirrored behavior. The green shaded region is centered around $I_{\text{bias}} \approx 17/2 \, \mu\text{A}$, where the system exhibits its most pronounced chaotic response. 
Notably, $ \bar{\beta}$ crosses the zero line at three distinct points within the green region, i.e., the 3WM nonlinearity alternates between positive and negative values multiple times. The frequent changes in the sign of $\bar{\beta}$ attest the complex dynamics of the system in this regime.
Additionally, both $\sigma_\beta$ and $\sigma_\gamma$ are particularly high within the green region, reflecting the enhanced variability and chaotic nature of the system. This strong nonlinear response is associated with the fact that $\varphi \sim \pi$, where the system is farthest from stable behavior.

These results support further the spectral analysis presented previously and the appearance of the 3WM idler peak for certain bias current values. The periodic changes in the coefficients, especially with varying $I_{\text{bias}}$, relate to the alternating influence of 3WM and 4WM processes, which contribute to the intricate nonlinear dynamics observed in the system.

\section{Effect of a non--sinusoidal CPR}
\label{Sec4}

In this section, we assume a JTWPA formed by rf--SQUIDs containing a \emph{transparency-dependent junction}, often embodied in a superconductor–normal conductor–superconductor (SNS) junction, showcasing an intrinsically non-sinusoidal CPR, like~\cite{Golubov2004, Beenakker1991, Beenakker1992}
\begin{equation}\label{nonsinusoidalCPR}
 I_{\tau}(\varphi) = \frac{\tau \sin \varphi}{2\sqrt{1 - \tau \sin^2 \varphi}},
\end{equation}
where $\tau \in [0, 1]$ is the \emph{transparency}. Its limit values lead to distinctive behaviors: for $\tau \to 1$, the junction becomes highly transparent, inducing a skewed CPR with $I_{\text{max}}= \max_{\varphi}[I_{\tau}(\varphi)] \to 1$, while for $\tau \to 0$, opacity prevails, yielding a more sinusoidal CPR but at the same time $I_{\text{max}} \to 0$.
In general, this type of CPR actually contains a sum of such contributions, one for each conduction channel, and with channel-dependent transparencies. Instead, here, as we are mainly interested in the effect produced by the anharmonicity of the CPR on the JTWPA behavior, we limit ourselves to considering only one contribution, as in Eq.~\eqref{nonsinusoidalCPR}. In this perspective, $\tau$ has the meaning of an \emph{effective transparency}, with the sole aim of capturing the skewness of the CPR. 
Finally, we mention that many experimental techniques for measuring the CPR of JJs exist~\cite{Waldram1975,VanHarlingen1995,Basset2014,Ginzburg2018}. 
%Comparison with a specific experimental setup would certainly involve a more complete description, with all necessary contributions into the CPR. 

\begin{figure*}[t!!]
\centering
\includegraphics[width=2\columnwidth]{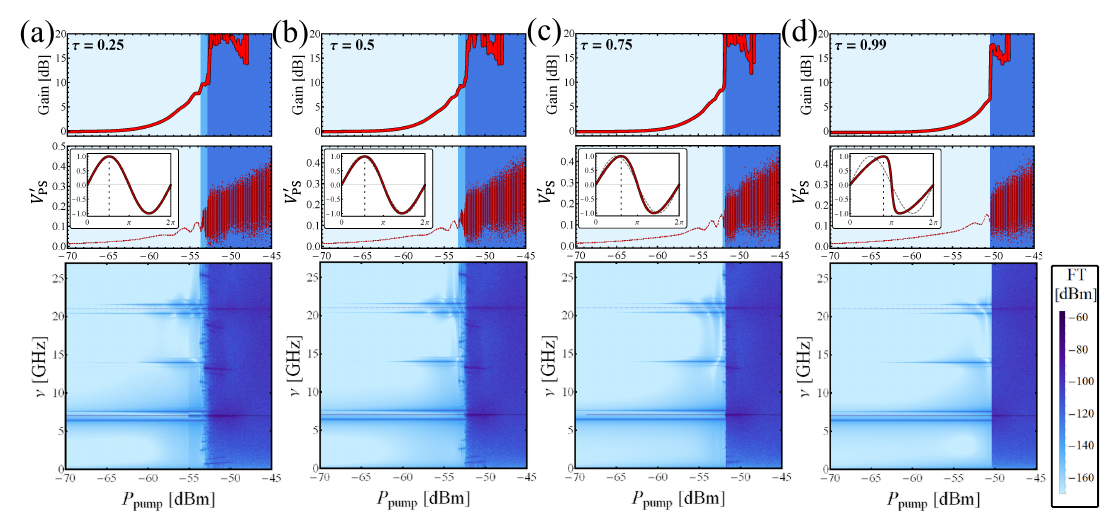}
\caption{Gain (top panel), PS (middle panel), and Fourier spectra of the output voltage (bottom panel) \emph{versus} the pump power level, $P_{\text{pump}}$, at different junction transparencies, i.e, $\tau=\{ 0.25, 0.5, 0.75, \text{ and } 0.99 \}$, see panels (a), (b), (c), and (d), respectively. In middle panels, the inset shows the CPR at the specific $\tau$ values considered and a sinusoidal profile, traced with a gray dashed curve. In top and middle panels, different response regimes are highlighted by regions shaded in different colors. In the density plots, the color intensity scale represents the amplitude of the spectral components whose frequency can be read on the bottom--left vertical axis. The other parameters are: $\nu_{\text{sign}}=6.42\;\text{GHz}$, $\nu_{\text{pump}}=7\;\text{GHz}$, $P_{\text{sign}}=-100\;\text{dBm}$, and $I_{\text{bias}}=0$.}
\label{Fig04}
\end{figure*}

Thinking about a specific experimental realisation, we must bear in mind that the anharmonic content of CPR can be determined not only by the nature and geometry of the junction, but also by temperature, transport parameters of the superconducting leads, and details of interfaces~\cite{Likharev1979,Golubov2004,Tafuri2019}. Indeed, as well as conventional SNS junctions, other concrete examples of systems show non-sinusoidal CPRs, e.g., based on topological insulators~\cite{Sochnikov2015,Kayyalha2020,Endres2023}, graphene~\cite{English2016,Nanda2017}, high-$T_c$ superconductors~\cite{Il'ichev1998,Il'ichev2001,Revin2018}, spatially
inhomogeneous ferromagnetic weak link~\cite{Golubov2002,Frolov2004}, nanobridge~\cite{Lindelof1981,Troeman2008,Wang2023} or nanowire~\cite{Spanton2017,Hart2019}, superconducting atomic contacts~\cite{DellaRocca2007}, or Cooper pair transistors~\cite{Paila2009}. Nonlinear CPRs occur even in conventional, i.e., Al--based, junctions~\cite{Willsch2024}, for possible CPR deviations due to the non-uniformity of the tunnel barrier~\cite{Aref2014,Zeng2015}. In general, the lower the temperature, the more skewed the CPRs are, thereby further supporting the need to explore divergences from pure sinusoidal behaviour, as JTWPAs generally operate at very low temperatures close to the quantum limit.
%Moreover, the skewness of the CPR can be even affected by voltage gating, temperature, or dimensions of the system. 
%Finally, we mention that many experimental techniques for measuring the CPR of JJs exist~\cite{Waldram1975,VanHarlingen1995,Basset2014,Ginzburg2018}. The most straightforward and effective way involves deducing the CPR's shape through the magnetic dependence of the critical current in a SQUID formed by two JJs with quite different critical currents.

Our focus now centers on the impact of transparency on JTWPA performance. Specifically, we narrow our attention to the non--sinusoidal nature of the CPR. To this aim, we express the Josephson current as the product of the critical current, $I_c$, and a dimensionless function $i(\varphi, \tau)\in[-1,+1]$ embodying the joint dependence on both the phase and the transparency, i.e., $I(\varphi) = I_c \cdot i(\varphi, \tau)$. The function $i(\varphi, \tau)$, which captures the skewness of the CPR, is the ratio between $I_{\tau}(\varphi)$ and its maximum value, that is
%\begin{equation}
%\varphi_{\text{max}} = \cos^{-1}\left(1 - 2\frac{1 - \sqrt{1 - \tau}}{\tau}\right),
%\end{equation}
%so that
%\begin{equation}
% I_{\text{max}}(\tau) = \frac{\tau}{\sqrt{2 - \tau + 2\sqrt{1 - \tau}}}.
%\end{equation}
%Thus, the dimensionless function, $i(\varphi, \tau)$, capturing the skewness of the CPR is:
\begin{equation}
% i(\varphi, \tau) &=& \frac{1}{I_{\text{max}}(\tau)} \, \frac{\tau \sin \varphi}{2\sqrt{1 - \tau \sin^2 \varphi}} \\
 i\left(\varphi, \tau\right) =  \frac{\sin \varphi\left(1 - \sqrt{1 - \tau}\right)}{2\sqrt{1 - \tau \sin^2 \varphi}}.
\end{equation}

\begin{figure}[b!!]
\centering
\includegraphics[width=0.85\columnwidth]{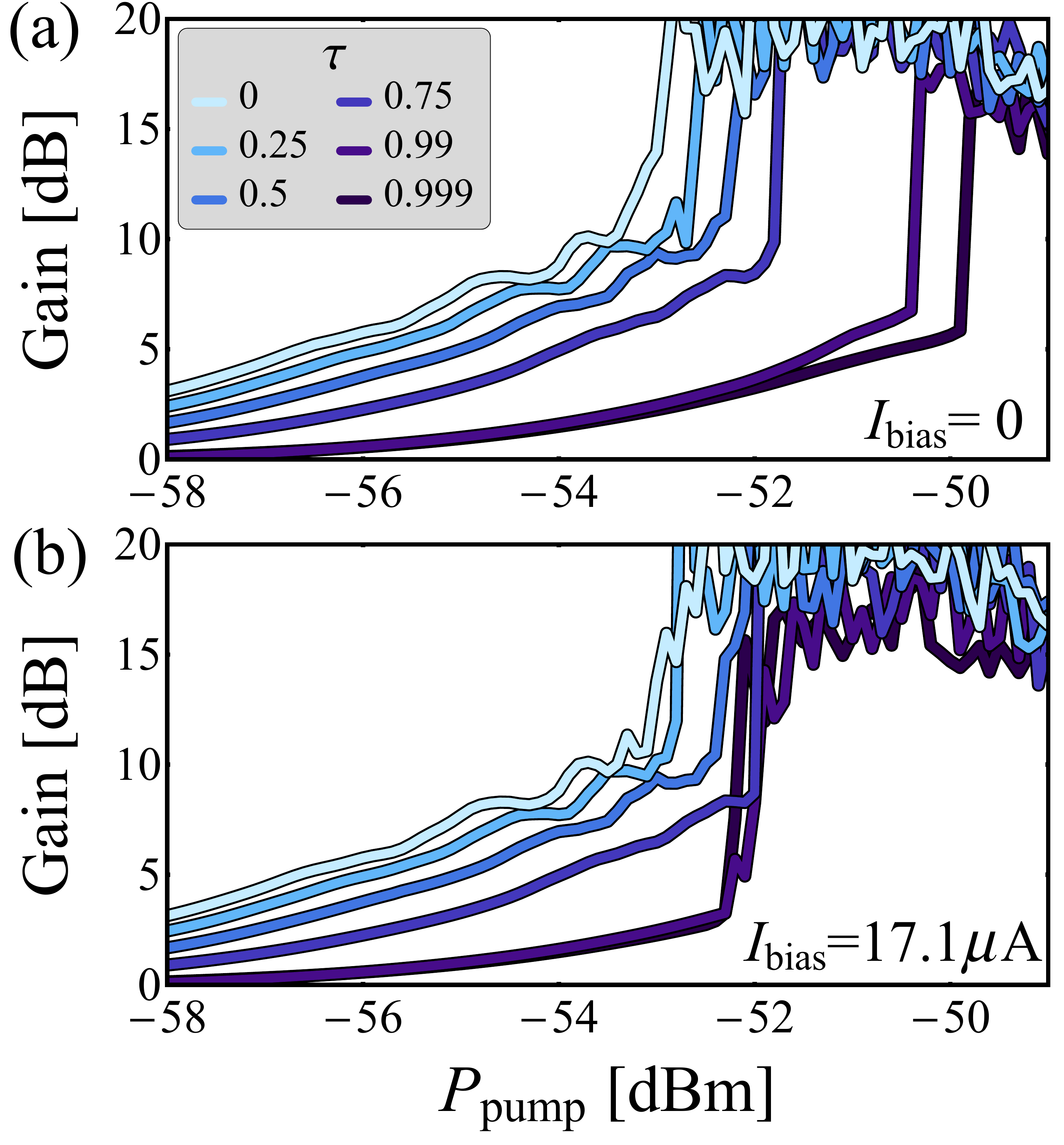}
\caption{(a) and (b), Gain \emph{versus} pump power level, $P_{\text{pump}}$, choosing different junction transparency, at $I_{\text{bias}}=0$ and $I_{\text{bias}}=17.1\;\mu\text{A}$, respectively. The other parameters are: $\nu_{\text{sign}}=6.42\;\text{GHz}$, $\nu_{\text{pump}}=7\;\text{GHz}$, and $P_{\text{sign}}=-100\;\text{dBm}$.}
\label{Fig05}
\end{figure}

In the following, we assume again a critical current equal to $I_c = 2\, \mu \text{A}$, just like in the previous section, and we adjust the junction transparency, $\tau$, i.e., the skewness of the CPR. In Fig.~\ref{Fig04}, we collect the gain (top panel), PS (middle panel), and Fourier spectra of the output voltage (bottom panel) \emph{versus} the pump power level, $P_{\text{pump}}$, at different $\tau=\{ 0.25, 0.5, 0.75, \text{ and } 0.99 \}$, see panels (a), (b), (c), and (d), respectively. The other parameters are: $\nu_{\text{sign}}=6.42\;\text{GHz}$, $\nu_{\text{pump}}=7\;\text{GHz}$, $P_{\text{sign}}=-100\;\text{dBm}$, and $I_{\text{bias}}=0$. In middle panels, the inset contains the CPR at the specific $\tau$ values considered and a sinusoidal profile, traced with a gray dashed curve.

\begin{figure*}[t!!]
\centering
\includegraphics[width=2\columnwidth]{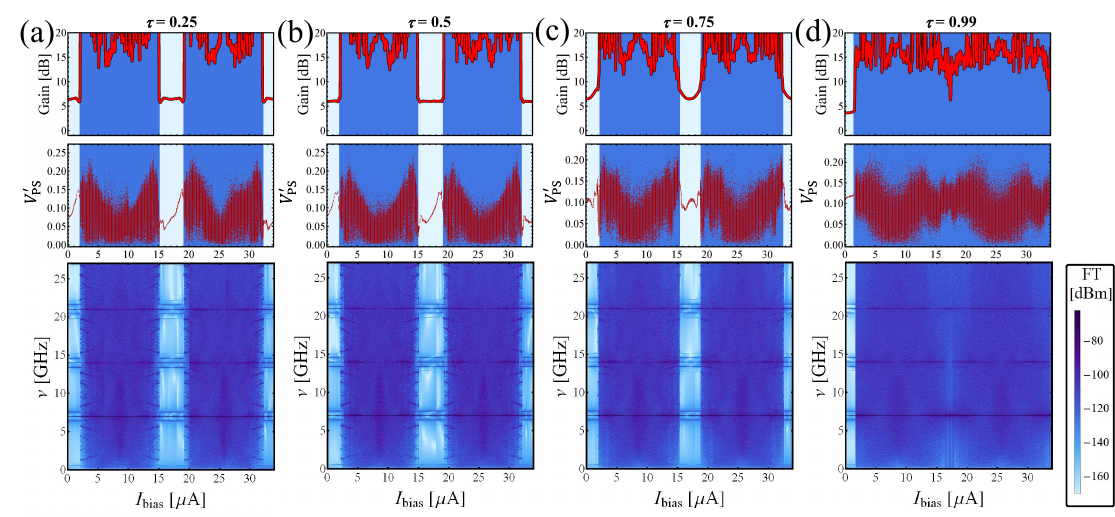}
\caption{Gain (top panel), PS (middle panel), and Fourier spectra of the output voltage (bottom panel) \emph{versus} the bias current, $I_{\text{bias}}$, at different junction transparencies, i.e, $\tau=\{ 0.25, 0.5, 0.75, \text{ and } 0.99 \}$, and pump intensities, i.e., $P_{\text{pump}}=\{-55, -54.5,-53,\text{ and }-52\}\; \text{dBm}$, see panels (a), (b), (c), and (d), respectively. In top and middle panels, different response regimes are highlighted by regions shaded in different colors. In the density plots, the color intensity scale represents the amplitude of the spectral components whose frequency can be read on the bottom--left vertical axis. The other parameters are: $\nu_{\text{sign}}=6.42\;\text{GHz}$, $\nu_{\text{pump}}=7\;\text{GHz}$, and $P_{\text{sign}}=-100\;\text{dBm}$.}
\label{Fig06}
\end{figure*}

Comparing results as the CPR skewness increases (i.e., $\tau\to1$) we see qualitatively similar behaviors, that is, an increasing trend as $P_{\text{pump}}$ enlarges, until a chaotic response is set. 
As $\tau$ increases, the region of $\varphi$ values around $\varphi=0$ in which the linear approximation of the CPR progressively enlarges; therefore, mixing processes, which need the nonlinearity of the CPR, become less efficient as $\tau\to 1$. This is clearly confirmed by comparying several gain curves for different values of $\tau$ in Fig.~\ref{Fig05}(a): the increase in the skewness of the CPR leads to a decrease in the maximum achievable gain, but on the other hand to a widening of the pump amplitude range giving a non--chaotic response.
Looking at the PS plots, we note also that the transition to the chaotic regime is gradually sharper as $\tau$ increases. This is also evident from the FT maps in the lower panels of this figure: in fact, the area of chaotic response, highlighted by the darker blue shades, for $\tau=0.99$ does not exhibit those patterns of more intense peaks at the stable/unstable transition, which are instead noticeable in the other cases (and that are increasingly more intense as $\tau$ decreases). 

There does not appear to be such a strong dependence on the frequency of the driving signal when changing $\tau$; in particular, even when choosing pump amplitudes just below the threshold beyond which a chaotic regime is established, a frequency sweep produces responses qualitatively similar to those shown in Fig.~\ref{Fig02}(b) (data not shown). 

In contrast, the situation is more interesting when considering the bias current effects at different $\tau$, see Fig.~\ref{Fig06}. Specifically, we explored the system response changing $I_{\text{bias}}$, setting a pump amplitude about a $1\;\text{dBm}$ below the chaos onset thresholds highlighted in Fig.~\ref{Fig04}. The increase of the anharmonic content of the CPR makes sharper the transition to a chaotic regime triggered by a bias current change. In particular, for $\tau= 0.25 \text{ and } 0.5$ we still note the structures of more intense peaks in the FT maps once the chaotic response is settled, see Figs.~\ref{Fig06}(a) and (b), respectively; conversely, when $\tau=0.75$ we notice that the distributions of marked spectral lines, prelude to the onset of chaos, are just barely appreciable, just to vanish completely when $\tau=0.99$, see Figs.~\ref{Fig06}(c) and (d), respectively. Furthermore, in the case of highly skewed CPR, i.e., $\tau=0.99$, we note the absence of those regions of $I_{\text{bias}}$ values with stable responses around $\sim17\mu\text{A}$ multiples. In other words, an high anharmonic CPR content seems to hinder the restoration of stable conditions once high bias currents induced a $2\pi$--phase rotation. 
Thus, we looked at the $P_{\text{pump}}$ dependence of the gain curves, at different values of $\tau$, setting this time $I_{\text{bias}}=17.1\;\mu\text{A}$, see Fig.~\ref{Fig05}(b). Comparing with the results in panel (a) obtained for $I_{\text{bias}}=0$, it can be seen that the threshold pump intensities, above which a chaotic response is triggered, tends to shift lower and lower as $\tau$ increases when a bias current $I_{\text{bias}}=17.1\;\mu\text{A}$ is applied: for instance, in the case of $\tau=0.99$ this threshold value is reduced by more than $2\;\text{dBm}$ compared to the case without current. 

\section{Conclusions} 
\label{Sec5}

In this work we have investigated the effects produced by variation of the input parameters in the response of a JTWPA formed by a sequence of cells, each containing an rf--SQUID and a ground capacity. In particular, we have characterized the parametric amplification process of a small--amplitude input signal, in the presence of a large-amplitude input tone. The Poincar\'e sections, together with the Fourier spectra analysis, reveal the intertwining between amplification process and the onset of chaos as system parameters are varied. In particular, we have discussed the effects of the pump intensity, the signal frequency, and the bias current, for circuit specifications in line with the device developed within the INFN DARTWARS collaboration~\cite{Pagano2022}. 

This study aims to improve the design of the future generation of JTWPAs, towards optimal performance and greater stability against chaotic disturbances, by deepening our understanding of their operational limits and capabilities.
This exploration not only contributes to the existing body of knowledge on the nonlinear dynamics of superconducting devices, but also underscores the importance of examining the broader spectrum of dynamic behaviors that can occur in complex quantum systems. 

The implications of chaos for JTWPA operation can be quite relevant, for it impacts amplifier performance and signal integrity. The transition to chaos leaves a spectral signature on the output voltage that correlates to the amplification gain of the input signal.
We construct Fourier maps, which serve as a guide to identify optimal operating conditions and to understand the limits of the amplification mechanism, before entering chaotic regimes.
At lower pump intensities, we observe a minimal signal response, indicating insufficient energy to trigger a fruitful parametric amplification processes. As the pump intensity increases, distinct spectral lines appear, corresponding to amplification and possibly mixing processes within the JTWPA. Notably, we recognize regions where the spectral lines broaden or split, which may suggest the onset of nonlinear phenomena or transitions to chaotic dynamics. 
The output signal's spectrum is quite sensitive to the bias current flowing through the transmission line. We observe regions where the spectrum is relatively undisturbed, with high gains, but as the bias current moves away from this region, dramatic changes in the output emerge, indicative of nonlinear and potentially chaotic behavior in the JTWPA. The periodicity, observed in the device response by changing the bias current, reflects the inherent phase matching condition between the two arms of the rf--SQUIDs. 

Our findings demonstrate that JTWPAs exhibit a sensitive dependence on drive parameters, for slight variations can cause a transition from predictable amplification to unpredictable chaos. This sensitivity underscores the critical importance of precision in parameter selection for optimal JTWPA operation.

The exploration has also included non--sinusoidal current--phase relations, offering a comprehensive view of JTWPAs crafted with diverse junction transparencies. We observe a device response quite sensitive to the shape of the Josephson current--phase relation. For example, increasing its skewness reduces the maximum achievable gains, but on the other hand increases the pump--signal intensity ranges within which the system behaviour is stable. Furthermore, the transition to the chaotic regime is much sharper for high skewness, but for very high values those windows of bias current values, at integer multiples of $\Phi_0/L_g$, within which the system response is stable are lost. We stress the importance of exploring deviations from pure sinusoidal behavior, since JTWPAs typically operate at very low temperatures near the quantum limit, where the current-phase relation tends to become increasingly skewed. Furthermore, at these low temperatures, the reduction in thermal noise, combined with the presence of broadband noise, can enhance chaotic dynamics and increase stochasticity, e.g., see Ref.~\cite{Pankratov2017}. These factors underscore the need for a thorough investigation of chaotic effects in JTWPAs to optimize their low--temperature performance.

In conclusion, our results are particularly relevant to the flourishing field of quantum technologies, where JTWPAs are poised to become indispensable tools for signal amplification at the quantum limit. 

\begin{acknowledgments}
This work was supported in part by the Italian Institute of Nuclear Physics (INFN) through the DARTWARS and QUB--IT Projects, in part by the European Union’s H2020--MSCA under Grant 101027746, in part by y the University of Salerno--Italy under Grants FRB21CAVAL, FRB22PAGAN, and FRB23BARON, and in part by PRIN 2022 PNRR Project QUESTIONs (Grant No. P2022KWFBH).
\end{acknowledgments}

\appendix

\section{The numerical approach}
\label{AppA}

In this appendix we present the numerical scheme used to get the solution of the setup in Fig.~\ref{Fig01}.

We start considering a generic mesh labeled with ``n''. The current balance at node \textbf{$n$} is
\begin{equation}\label{Eq01}
I_n=\dot{q}_n+I_{n+1}
\end{equation}
while the current balance at node \textbf{$n_{JL}$} is 
\begin{equation}\label{Eq02}
I_n=I_{J,n}+I_{L,n}
\end{equation}
where
\begin{equation}\label{Eq03}
I_{J,n}=C_{J,n}\frac{\hbar}{2e}\frac{d^2\varphi_{n}}{dt^2}+\frac{1}{R_{J,n}}\frac{\hbar}{2e}\frac{d\varphi_{n}}{dt}+I_{c,n}\sin\varphi_{n} 
\end{equation}
and
\begin{equation}\label{Eq04}
I_{L,n}=\frac{1}{L_{g,n}}\frac{\hbar}{2e}\varphi_{n}.
\end{equation}
The voltage drop in the mesh ``n'', with $\text{n}\in[1,...,N]$, is 
\begin{equation}\label{Eq05}
-\frac{q_{n-1}}{C_{n-1}}+\frac{\hbar}{2e}\frac{d\varphi_{n}}{dt}+\frac{q_{n}}{C_{n}}=0
\end{equation}
from which
\begin{equation}\label{Eq06}
-\frac{\dot{q}_{n-1}}{C_{n-1}}+\frac{\hbar}{2e}\frac{d^2\varphi_{n}}{dt^2}+\frac{\dot{q}_{n}}{C_{n}}=0.
\end{equation}
Inserting Eq.~\eqref{Eq01} into Eq.~\eqref{Eq06} one obtains
\begin{eqnarray}\label{Eq07}\nonumber
-\frac{I_{n-1}-I_n}{C_{n-1}}+\frac{\hbar}{2e}\frac{d^2\varphi_{n}}{dt^2}+\frac{I_n-I_{n+1}}{C_{n}}&=&0\\\label{Eq08}\nonumber
-\frac{I_{n-1}}{C_{n-1}}-\frac{I_{n+1}}{C_{n}}+I_{n}\left ( \frac{1}{C_{n-1}}+\frac{1}{C_{n}} \right )+\frac{\hbar}{2e}\frac{d^2\varphi_{n}}{dt^2}&=&0.
\end{eqnarray}
Defining the quantities (a tilde over a label indicates the normalization to $\frac{2e}{\hbar}$)
\begin{eqnarray}\label{Eq09}
\widetilde{C}_{J,n}=C_{J,n}\frac{\hbar}{2e}\qquad&&\qquad\widetilde{R}_{J,n}^{-1}=R_{J,n}^{-1}\frac{\hbar}{2e}\\
\widetilde{L}_{g,n}^{-1}=L_{g,n}^{-1}\frac{\hbar}{2e}\qquad&&\qquad\widetilde{C}_{n}=C_{n}\frac{\hbar}{2e}
\end{eqnarray}
and inserting Eqs.~\eqref{Eq03}-\eqref{Eq04} in Eqs.~\eqref{Eq02}, one obtains
\begin{equation}\label{Eq10}
I_{n}=\widetilde{C}_{J,n}\frac{d^2\varphi_{n}}{dt^2}+\frac{1}{\widetilde{R}_{J,n}}\frac{d\varphi_{n}}{dt}+I_{c,n}\sin\varphi_{n} +\frac{1}{\widetilde{L}_{g,n}}\varphi_{n}
\end{equation}
and Eq.~\eqref{Eq08}, defining $C_{n}^{-}=\frac{C_{n}}{C_{n-1}}$, can be recast as
\begin{equation}\label{Eq11}
-C_{n}^{-}I_{n-1}+\left ( 1+C_{n}^{-} \right )I_{n}+\widetilde{C}_{n}\frac{d^2\varphi_{n}}{dt^2}-I_{n+1}=0.
\end{equation}
Inserting Eq.~\eqref{Eq10} in Eqs.~\eqref{Eq11}
\begin{widetext}
\begin{eqnarray}\label{Eq12}
-C_{n}^{-}&&\left [ \widetilde{C}_{J,n-1}\frac{d^2\varphi_{n-1}}{dt^2}+\frac{1}{\widetilde{R}_{J,n-1}}\frac{d\varphi_{n-1}}{dt}+I_{c,n-1}\sin\varphi_{n-1} +\frac{1}{\widetilde{L}_{g,n-1}}\varphi_{n-1} \right ]\\\label{Eq13}
+\left ( 1+C_{n}^{-} \right )&&\left [ \left ( \widetilde{C}_{J,n} +\frac{\widetilde{C}_{n}}{1+C_{n}^{-}}\right )\frac{d^2\varphi_{n}}{dt^2}+\frac{1}{\widetilde{R}_{J,n}}\frac{d\varphi_{n}}{dt}+I_{c,n}\sin\varphi_{n} +\frac{1}{\widetilde{L}_{g,n}}\varphi_{n} \right ]\\\label{Eq14}
-&&\left [ \widetilde{C}_{J,n+1}\frac{d^2\varphi_{n+1}}{dt^2}+\frac{1}{\widetilde{R}_{J,n+1}}\frac{d\varphi_{n+1}}{dt}+I_{c,n+1}\sin\varphi_{n+1} +\frac{1}{\widetilde{L}_{g,n+1}}\varphi_{n+1} \right ]=0
\end{eqnarray}
\end{widetext}

\emph{Discretization} $-$
For the numerical integration of previous equations, the time was divided into many short time intervals $k=\Delta t=t_{max}/M$, where $t_{max}$ and $M$ are the observation time and the number of intervals, respectively. The partial derivatives are approximated using the Euler formalism. 
The phase $\varphi_n(t)$ is labeled by $\varphi_n^m=\varphi_n(mk)$, where $n$ and $m$ are the discrete mesh and time indexes, respectively, that is $\varphi_{n\;\to\; \text{mesh index}}^{m\; \to\; \text{time index}}\;\;.$
The partial derivatives can be expressed as
\begin{eqnarray}\label{Eq15}
&&\frac{\partial \varphi_n}{\partial t}\simeq\frac{\varphi_n^{m+1}-\varphi_n^{m-1}}{2k}\\
&&\frac{\partial^2 \varphi_n}{\partial t^2}\simeq\frac{\varphi^{m+1}_n-2\varphi_{n}^m+\varphi^{m-1}_n}{k^2}.
\end{eqnarray}
In this way, Eq.~\eqref{Eq10} can be rewritten as
\begin{widetext}
\begin{eqnarray}\label{Eq16}
I_{n}&\simeq&\frac{\widetilde{C}_{J,n}}{k^2}\left ( \varphi^{m+1}_n-2\varphi_{n}^m+\varphi^{m-1}_n \right )+\frac{1}{2\widetilde{R}_{J,n}k}\left (\varphi_n^{m+1}-\varphi_n^{m-1} \right )+I_{c,n}\sin\varphi_n^m +\frac{\varphi_n^m}{\widetilde{L}_{g,n}}\\\label{Eq17}
I_{n}&\simeq&\varphi^{m+1}_n\left ( \frac{\widetilde{C}_{J,n}}{k^2}+\frac{1}{2\widetilde{R}_{J,n}k} \right )+
\left ( -\frac{2\widetilde{C}_{J,n}}{k^2}\varphi_{n}^m+I_{c,n}\sin\varphi_n^m +\frac{\varphi_n^m}{\widetilde{L}_{g,n}} \right )
+\varphi^{m-1}_n\left ( \frac{\widetilde{C}_{J,n}}{k^2}-\frac{1}{2\widetilde{R}_{J,n}k} \right ),\qquad
\end{eqnarray}
\end{widetext}
which becomes
\begin{equation}\label{Eq19}
I_{n}\simeq\alpha_{n}^{+}\varphi^{m+1}_{n}+f_{n}^{m}+\alpha_{n}^{-}\varphi^{m-1}_{n}
\end{equation} 
after defining the quantities
\begin{eqnarray}\label{Eq18}
&&\alpha_n^{\pm}=\frac{\widetilde{C}_{J,n}}{k^2}\pm\frac{1}{2\widetilde{R}_{J,n}k}\\
&&f_n^m=\left (\frac{1}{\widetilde{L}_{g,n}}-\frac{2\widetilde{C}_{J,n}}{k^2}\right )\varphi_{n}^m+I_{c,n}\sin\varphi_n^m.
\end{eqnarray}
These coefficients take a slightly different expression when considering Eq.~\eqref{Eq13}, that is
\begin{equation}
\widetilde{\alpha}_n^{\pm}=\frac{\widetilde{C}_{\text{``n''}}}{k^2}\pm\frac{1}{2\widetilde{R}_{J,n}k}
\end{equation}
\begin{equation}
\widetilde{f}_n^m=\left (\frac{1}{\widetilde{L}_{g,n}}-\frac{2\widetilde{C}_{\text{``n''}}}{k^2}\right )\varphi_{n}^m+I_{c,n}\sin\varphi_n^m,
\end{equation}
%
%where we have defined the capacitance 
\begin{equation}
\widetilde{C}_{\text{``n''}}=\left ( \widetilde{C}_{J,n} +\frac{\widetilde{C}_{n}}{1+C_{n}^{-}}\right )=\frac{\hbar}{2e}\left ( C_{J,n} +\frac{C_{n-1}C_{n}}{C_{n-1}+C_{n}}\right ).
\end{equation}
Similarly, after discretization of the terms in square brackets in Eqs.~\eqref{Eq12}-\eqref{Eq14}, one obtains:
\begin{widetext}
\begin{eqnarray}\label{Eq20}\nonumber
&&-C_{n}^{-}\alpha_{n-1}^{+}\varphi^{m+1}_{n-1}+\left ( 1+C_{n}^{-} \right )\widetilde{\alpha}_{n}^{+}\varphi^{m+1}_{n}-\alpha_{n+1}^{+}\varphi^{m+1}_{n+1}=\\
&&C_{n}^{-}f_{n-1}^{m}-\left ( 1+C_{n}^{-} \right )\widetilde{f}_{n}^{m}+f_{n+1}^{m}
+C_{n}^{-}\alpha_{n-1}^{-}\varphi^{m-1}_{n-1}-\left ( 1+C_{n}^{-} \right )\widetilde{\alpha}_{n}^{+}\varphi^{m-1}_{n}+\alpha_{n+1}^{-}\varphi^{m-1}_{n+1}
\end{eqnarray}
that becomes
\begin{equation}\label{Eq20}
a_{n,1}\varphi^{m+1}_{n-1}+a_{n,2}\varphi^{m+1}_{n}+a_{n,3}\varphi^{m+1}_{n+1}=
b_{n,1}f^{m}_{n-1}+b_{n,2}\widetilde{f}^{m}_{n}+b_{n,3}f^{m}_{n+1}+
c_{n,1}\varphi^{m-1}_{n-1}+c_{n,2}\varphi^{m-1}_{n}+c_{n,3}\varphi^{m-1}_{n+1}
\end{equation}
defining the coefficients:
\begin{equation}\label{Tab01}
\begin{tabular}{lll}
$a_{n,1}=-C_{n}^- \alpha_{n-1}^{+}$ & \qquad$a_{n,2}=\left ( 1+C_{n}^{-} \right ) \widetilde{\alpha}_{n}^{+}$ &\qquad$a_{n,3}= -\alpha_{n+1}^{+}$ \\
$b_{n,1} =C_{n}^{-}$&\qquad$b_{n,2} =-\left ( 1+C_{n}^{-} \right )$ &\qquad $b_{n,3}=1$ \\
$c_{n,1}=C_{n}^{-} \alpha_{n-1}^{-}$&\qquad$c_{n,2}=-\left ( 1+C_{n}^{-} \right ) \widetilde{\alpha}_{n}^{-}$ &\qquad $c_{n,3}= \alpha_{n+1}^{-}$
\end{tabular}
\end{equation}
\end{widetext}

%
%\begin{figure*}[t!!]
%\centering
%\includegraphics[width=0.45\textwidth]{Device-bc_Left.pdf}\hspace{1.475cm}
%\includegraphics[width=0.45\textwidth]{Device-bc_Rigth.pdf}
%\caption{Boundary conditions}
%\label{Fig01}
%\end{figure*}
%

\emph{Left Boundary Conditions} $-$
Let's first consider the left side of the circuit including the voltage generator $V_i(t)$. 

The current balances and voltage drops at the leftmost mesh in left panel of Fig.~\ref{Fig01} gives
\begin{eqnarray}\label{Eq21}
&&V_i=I_iR_i+\frac{q_0}{C_i}\qquad\qquad I_i+I_{\text{bias}}=\dot{q}_0+I_{1}\\\label{Eq22}\nonumber
&&I_1=C_{J,1}\frac{\hbar}{2e}\frac{d^2\varphi_{1}}{dt^2}+\frac{1}{R_{J,1}}\frac{\hbar}{2e}\frac{d\varphi_{1}}{dt}+I_{c,1}\sin\varphi_{1}+\frac{1}{\widetilde{L}_{g,1}}\varphi_{1} \\\label{Eq23}\nonumber
%-\frac{q_{0}}{C_{0}}+\frac{\hbar}{2e}\frac{d\varphi_{1}}{dt}+\frac{q_{1}}{C_{1}}&=&0.
\end{eqnarray}
Combining these equations, one obtains
\begin{equation}\label{Eq24}
\dot{I}_i=\frac{\dot{V}_i}{R_i}-\frac{\dot{q}_0}{R_iC_i}=\frac{\dot{V}_i}{R_i}-\frac{I_i+I_{\text{bias}}-I_{1}}{R_iC_i}
\end{equation}
and defining $\omega_i=1/(R_iC_i)$ we achieve the equation
\begin{widetext}
\begin{equation}
\dot{I}_i
=-\omega_iI_i+\left (\frac{\dot{V}_i}{R_i}-\omega_iI_{\text{bias}}\right )+\omega_i\left (\widetilde{C}_{J,1}\frac{d^2\varphi_{1}}{dt^2}+\frac{1}{\widetilde{R}_{J,1}}\frac{d\varphi_{1}}{dt}+I_{c,1}\sin\varphi_{1}+\frac{1}{\widetilde{L}_{g,1}}\varphi_{1}\right ).
\end{equation}
After discretization, this equation becomes
\begin{eqnarray}\label{Eq25}\nonumber
\frac{I_i^{m+1}-I_i^{m-1}}{2k\omega_i}=&&-I_i^m+\left (\frac{\dot{V}_i^m}{\omega_iR_i}-I_{\text{bias}}\right )+
\varphi^{m+1}_1\left ( \frac{\widetilde{C}_{J,1}}{k^2}+\frac{1}{2\widetilde{R}_{J,1}k} \right )\\
&&+\left ( -\frac{2\widetilde{C}_{J,1}}{k^2}\varphi_{1}^m+I_{c,1}\sin\varphi_1^m +\frac{\varphi_1^m}{\widetilde{L}_{g,1}} \right )
+\varphi^{m-1}_1\left ( \frac{\widetilde{C}_{J,1}}{k^2}-\frac{1}{2\widetilde{R}_{J,1}k} \right )
\end{eqnarray}
so that
\begin{eqnarray}\label{Eq26}\nonumber
\frac{I_i^{m+1}}{2k\omega_i}-\varphi^{m+1}_1\left ( \frac{\widetilde{C}_{J,1}}{k^2}+\frac{1}{2\widetilde{R}_{J,1}k} \right )=&&-I_i^m+\left (\frac{\dot{V}_i^m}{\omega_iR_i}-I_{\text{bias}}\right )+\left [\left (\frac{1}{\widetilde{L}_{g,1}} -\frac{2\widetilde{C}_{J,1}}{k^2}\right )\varphi_{1}^m+I_{c,1}\sin\varphi_1^m \right ]\\
&&+\frac{I_i^{m-1}}{2k\omega_i}+\varphi^{m-1}_1\left ( \frac{\widetilde{C}_{J,1}}{k^2}-\frac{1}{2\widetilde{R}_{J,1}k} \right ).
\end{eqnarray}
and
\begin{equation}\label{Eq26}
\frac{I_i^{m+1}}{2k\omega_i}-\varphi^{m+1}_1\alpha_{1}^{+}=-I_i^m+f_1^m+\frac{I_i^{m-1}}{2k\omega_i}+\varphi^{m-1}_1\alpha_{1}^{-}+\left (C_i\dot{V}_i^m-I_{\text{bias}}\right ),
\end{equation}
and finally
\begin{equation}
I_i^{m+1}=I_i^{m-1}+\left [ \varphi^{m+1}_1\alpha_{1}^{+}-I_i^m+f_1^m+\varphi^{m-1}_1\alpha_{1}^{-}+\left (C_i\dot{V}_i^m-I_{\text{bias}}\right ) \right ]\left ( 2k\omega_i \right ).
\end{equation}
\end{widetext}

The current balances at nodes and voltage drops at the mesh labeled with ``1'' in the left panel of Fig.~\ref{Fig01} gives
\begin{eqnarray}
&I_i+I_{\text{bias}}-\dot{q}_0-I_{1}=0\\
&I_1-I_2-\dot{q_0}=0\\
&-\frac{\dot{q}_{0}}{C_{0}}+\frac{\hbar}{2e}\frac{d^2\varphi_{1}}{dt^2}+\frac{\dot{q}_{1}}{C_{1}}=0
\end{eqnarray}
from which
\begin{equation}
-\frac{(I_i+I_{\text{bias}})-I_{1}}{C_{0}}+\frac{\hbar}{2e}\frac{d^2\varphi_{1}}{dt^2}+\frac{\dot{q}_{1}}{C_{1}}=0.
\end{equation}
In other words, we are replacing $I_{n-1}$ with $(I_i+I_{\text{bias}})$ in Eq.~\eqref{Eq07}, so that
\begin{equation}
-C_{1}^{-}(I_i+I_{\text{bias}})+\left ( 1+C_{1}^{-} \right )I_{1}+\widetilde{C}_{1}\frac{d^2\varphi_{1}}{dt^2}-I_{2}=0,
\end{equation}
which, making all terms explicit, becomes
\begin{widetext}
\begin{eqnarray}\nonumber
-C_{1}^{-}\left [ I_i+I_{\text{bias}} \right ]
+\left ( 1+C_{1}^{-} \right )&&\left [ \left ( \widetilde{C}_{J,1} +\frac{\widetilde{C}_{1}}{1+C_{1}^{-}}\right )\frac{d^2\varphi_{1}}{dt^2}+\frac{1}{\widetilde{R}_{J,1}}\frac{d\varphi_{1}}{dt}+I_{c,1}\sin\varphi_{1} +\frac{1}{\widetilde{L}_{g,1}}\varphi_{1} \right ]\\
-&&\left [ \widetilde{C}_{J,2}\frac{d^2\varphi_{2}}{dt^2}+\frac{1}{\widetilde{R}_{J,2}}\frac{d\varphi_{2}}{dt}+I_{c,2}\sin\varphi_{2} +\frac{1}{\widetilde{L}_{g,2}}\varphi_{2} \right ]=0.
\end{eqnarray}
After discretization, we obtain
\begin{eqnarray}
-C_{1}^{-}\left [ I_i^m+I_{\text{bias}} \right ]
+\left ( 1+C_{1}^{-} \right )\left [ \widetilde{\alpha}_{1}^{+}\varphi^{m+1}_{1}+\widetilde{f}_{1}^{m}+\widetilde{\alpha}_{1}^{-}\varphi^{m-1}_{1} \right ]
-\left [ \alpha_{2}^{+}\varphi^{m+1}_{2}+f_{2}^{m}+\alpha_{2}^{-}\varphi^{m-1}_{2} \right ]=0.
\end{eqnarray}
By slightly manipulating this equation, one obtains
\begin{eqnarray}
\left ( 1+C_{1}^{-} \right )\widetilde{\alpha}_{1}^{+}\varphi^{m+1}_{1}-\alpha_{2}^{+}\varphi^{m+1}_{2}=C_{1}^{-} I_i^m
-\left ( 1+C_{1}^{-} \right )\widetilde{f}_{1}^{m}+f_{2}^{m}
-\left ( 1+C_{1}^{-} \right )\widetilde{\alpha}_{1}^{+}\varphi^{m-1}_{1}+\alpha_{2}^{-}\varphi^{m-1}_{2}+C_{1}^{-}I_{\text{bias}} 
\end{eqnarray}
that can be recast in a compact form as
\begin{equation}\label{bcl}
a_{1,2}\varphi^{m+1}_{1}+a_{1,3}\varphi^{m+1}_{2}=
b_{1,1}I_i^m+b_{1,2}\widetilde{f}^{m}_{1}+b_{1,3}f^{m}_{2}+
c_{1,2}\varphi^{m-1}_{1}+c_{1,3}\varphi^{m-1}_{2}+C_{1}^{-}I_{\text{bias}} 
\end{equation}
by defining the coefficients
\begin{equation}\label{Tab02}
\begin{tabular}{lll}
$a_{1,1}=0$ & \qquad$a_{1,2}=\left ( 1+C_{1}^{-} \right ) \widetilde{\alpha}_{1}^{+}$ &\qquad$a_{1,3}= -\alpha_{2}^{+}$ \\
$b_{1,1} =C_{1}^{-}$&\qquad$b_{1,2} =-\left ( 1+C_{1}^{-} \right )$ &\qquad $b_{1,3}=1$ \\
$c_{1,1}=0$&\qquad$c_{1,2}=-\left ( 1+C_{1}^{-} \right ) \widetilde{\alpha}_{1}^{-}$ &\qquad $c_{1,3}= \alpha_{2}^{-}$.
\end{tabular}
\end{equation}
\end{widetext}
Comparing with the coefficients in Eq.~\eqref{Tab01}, it means to impose $\alpha_{0}^{\pm}=0$. 

\emph{Right Boundary Conditions} $-$
Let's look now at the right side of the circuit, including $R_\ell(t)$ and $C_\ell(t)$.
The current balances at nodes and voltage drops at the rightmost mesh of Fig.~\ref{Fig01} gives
\begin{eqnarray}\label{Eq29}
I_N&=&I_{\text{bias}}+\dot{q}_N+\dot{q}_\ell\qquad,\qquad \frac{q_N}{C_N}=\frac{q_\ell}{C_\ell}+\dot{q}_\ell R_\ell\\\nonumber
I_N&=&\widetilde{C}_{J,N}\frac{d^2\varphi_{N}}{dt^2}+\frac{1}{\widetilde{R}_{J,N}}\frac{d\varphi_{N}}{dt}+I_{c,N}\sin\varphi_{N}+\frac{1}{\widetilde{L}_{g,n}}\varphi_{N}. \nonumber
%\frac{q_N}{C_N}&=&\frac{q_\ell}{C_\ell}+\dot{q}_\ell R_\ell
\end{eqnarray}
Through the time derivative of Eq.~\eqref{Eq29}, the following first-order differential equation for $I_\ell=\dot{q}_\ell$ is obtained
\begin{widetext}
\begin{equation}\label{Eq32}
C_N R_\ell\,\dot{I}_\ell =-\left (1+\frac{C_N}{C_\ell}\right )I_\ell-I_{\text{bias}}+\left (\widetilde{C}_{J,N}\frac{d^2\varphi_{N}}{dt^2}+\frac{1}{\widetilde{R}_{J,N}}\frac{d\varphi_{N}}{dt}+I_{c,N}\sin\varphi_{N}+\frac{1}{\widetilde{L}_{g,n}}\varphi_{N}\right ),
\end{equation}
from which, after discretization, one obtains
\begin{eqnarray}\label{Eq33}\nonumber
C_N R_\ell \frac{I_\ell^{m+1}-I_\ell^{m-1}}{2k}&=&-\left (1+\frac{C_N}{C_\ell}\right )I^m_\ell-I_{\text{bias}}+
\varphi^{m+1}_N\left ( \frac{\widetilde{C}_{J,N}}{k^2}+\frac{1}{2\widetilde{R}_{J,N}k} \right )\\
&&+\left ( -\frac{2\widetilde{C}_{J,N}}{k^2}\varphi_{N}^m+I_{c,N}\sin\varphi_N^m +\frac{\varphi_N^m}{\widetilde{L}_{g,n}} \right )
+\varphi^{m-1}_N\left ( \frac{\widetilde{C}_{J,N}}{k^2}-\frac{1}{2\widetilde{R}_{J,N}k} \right ),
\end{eqnarray}
that can be recast in
\begin{eqnarray}\nonumber
-\varphi^{m+1}_N\left ( \frac{\widetilde{C}_{J,N}}{k^2}+\frac{1}{2\widetilde{R}_{J,N}k} \right )+ \frac{C_N R_\ell}{2k}I_\ell^{m+1}&=&f_N^m-\left (1+\frac{C_N}{C_\ell}\right )I^m_\ell\\\label{Eq34}
 &+&\varphi^{m-1}_N\left ( \frac{\widetilde{C}_{J,N}}{k^2}-\frac{1}{2\widetilde{R}_{J,N}k} \right )+\frac{C_N R_\ell}{2k}I_\ell^{m-1}-I_{\text{bias}}.
\end{eqnarray}
\begin{eqnarray}\nonumber
 I_\ell^{m+1}=I_\ell^{m-1}+\Bigg [f_N^m-\left (1+\frac{C_N}{C_\ell}\right )I^m_\ell
 +\varphi^{m+1}_N\alpha_{N}^{+}+\varphi^{m-1}_N\alpha_{N}^{-}-I_{\text{bias}}\Bigg ]\frac{2k}{C_N R_\ell}.
\end{eqnarray}
\end{widetext}

The output voltage is finally equal to
\begin{equation}
V_{out}^{m+1}=I^{m+1}_\ell R_\ell.
\end{equation}

The balances of the currents at nodes and voltage drops at the \textbf{mesh labeled with ``N''} in the right panel of Fig.~\ref{Fig01} gives
\begin{eqnarray}
I_{N}-(I_\ell+I_{\text{bias}})&=&\dot{q}_N\\
I_{N-1}-I_N&=&\dot{q_{N-1}}\\
-\frac{\dot{q}_{N-1}}{C_{N-1}}+\frac{\hbar}{2e}\frac{d^2\varphi_{N}}{dt^2}+\frac{\dot{q}_{N}}{C_{N}}&=&0 
\end{eqnarray}
from which
\begin{eqnarray}\nonumber
-\frac{I_{N-1}-I_N}{C_{N-1}}+\frac{\hbar}{2e}\frac{d^2\varphi_{N}}{dt^2}+\frac{I_{N}-(I_\ell+I_{\text{bias}})}{C_{N}}&=&0.\\\nonumber
-C_{N}^{-}I_{N-1}+\left ( 1+C_{N}^{-} \right )I_{N}+\widetilde{C}_{N}\frac{d^2\varphi_{N}}{dt^2}-(I_\ell+I_{\text{bias}})&=&0.
\end{eqnarray}
that can be recast as
\begin{widetext}
\begin{eqnarray}
-C_{N}^{-}&&\left [ \widetilde{C}_{J,N-1}\frac{d^2\varphi_{N-1}}{dt^2}+\frac{1}{\widetilde{R}_{J,N-1}}\frac{d\varphi_{N-1}}{dt}+I_{c,N-1}\sin\varphi_{N-1} +\frac{1}{\widetilde{L}_{g,N-1}}\varphi_{N-1} \right ]\\
+\left ( 1+C_{N}^{-} \right )&&\left [ \left ( \widetilde{C}_{J,N} +\frac{\widetilde{C}_{N}}{1+C_{N}^{-}}\right )\frac{d^2\varphi_{N}}{dt^2}+\frac{1}{\widetilde{R}_{J,N}}\frac{d\varphi_{N}}{dt}+I_{c,N}\sin\varphi_{N} +\frac{1}{\widetilde{L}_{g,n}}\varphi_{N} \right ]-(I_\ell+I_{\text{bias}})=0
\end{eqnarray}
or alternatively as
\begin{eqnarray}
-C_{n}^{-}\left [ \alpha_{n-1}^{+}\varphi^{m+1}_{n-1}+f_{n-1}^{m}+\alpha_{n-1}^{-}\varphi^{m-1}_{n-1} \right ]
+\left ( 1+C_{n}^{-} \right )\left [ \widetilde{\alpha}_{n}^{+}\varphi^{m+1}_{n}+\widetilde{f}_{n}^{m}+\widetilde{\alpha}_{n}^{-}\varphi^{m-1}_{n} \right ]
-(I^m_\ell+I_{\text{bias}})=0.
\end{eqnarray}
and collecting the terms appropriately, we obtain
\begin{eqnarray}\nonumber
-C_{N}^{-}\alpha_{N-1}^{+}\varphi^{m+1}_{N-1}\!+\!\left ( 1+C_{N}^{-} \right )\widetilde{\alpha}_{N}^{+}\varphi^{m+1}_{N}\!=\!C_{N}^{-}f_{N-1}^{m}\!-\!\left ( 1+C_{N}^{-} \right )\widetilde{f}_{N}^{m}\!+\!I_\ell^m
\!+\!C_{N}^{-}\alpha_{N-1}^{-}\varphi^{m-1}_{N-1}\!-\!\left ( 1+C_{N}^{-} \right )\widetilde{\alpha}_{N}^{+}\varphi^{m-1}_{N}\!+\!I_{\text{bias}}
\end{eqnarray}
which can be written as
\begin{equation}\label{bcr}
a_{N,1}\varphi^{m+1}_{N-1}+a_{N,2}\varphi^{m+1}_{N}=b_{N,1}f^{m}_{N-1}+b_{N,2}\widetilde{f}^{m}_{N}+b_{n,3}I^{m}_\ell+
c_{N,1}\varphi^{m-1}_{N-1}+c_{N,2}\varphi^{m-1}_{N}+I_{\text{bias}}
\end{equation}
where
\begin{equation}
\begin{tabular}{lll}
$a_{N,1}=-C_{N}^- \alpha_{N-1}^{+}$ & \qquad$a_{N,2}=\left ( 1+C_{N}^{-} \right ) \widetilde{\alpha}_{N}^{+}$ &\qquad$a_{N,3}= 0$ \\
$b_{N,1} =C_{N}^{-}$&\qquad$b_{N,2} =-\left ( 1+C_{N}^{-} \right )$ &\qquad $b_{N,3}=1$ \\
$c_{N,1}=C_{N}^{-} \alpha_{N-1}^{-}$&\qquad$c_{N,2}=-\left ( 1+C_{N}^{-} \right ) \widetilde{\alpha}_{N}^{-}$ &\qquad $c_{n,3}=0$
\end{tabular}
\end{equation}
\end{widetext}
In other words, we are imposing $\alpha_{N+1}^{\pm}=0$ in Eq.~\eqref{Tab01}.

\emph{Matrix Representation of the System of Differential Equations} $-$
The matrix representing Eqs.~\eqref{Eq20}-\eqref{bcl}-\eqref{bcr} has non-zero elements only in the main, upper minor, and lower minor diagonals.
This tridiagonal matrix has a form given by:
\begin{widetext}
\begin{equation}
\left\|
\begin{array}
[c]{ll}%
a_{1,2}\hspace{0.5cm}a_{1,3} \hspace{0.5cm}0\hspace{1.35cm}\ldots\hspace{1cm}0\\
a_{2,1}\hspace{0.5cm}a_{2,2}\hspace{0.5cm}a_{2,3}\hspace{0.975cm}\ldots\hspace{0.99cm}0\\
\;\vdots\hspace{0.9cm}\ddots\hspace{0.6cm}\ddots\hspace{1cm}\ddots\hspace{1.075cm}\vdots\\
\;0\hspace{0.85cm}\ldots\hspace{0.5cm}a_{N-1,1}\hspace{0.32cm}a_{N-1,2}\hspace{0.6cm}a_{N-1,3}\\
\;0\hspace{0.85cm}\ldots\hspace{0.5cm}\;0\;\hspace{0.975cm}a_{N,1}\hspace{0.975cm}a_{N,2}
\;
\end{array}
\right\|
\left\|
\begin{array}
[c]{ll}%
\varphi_1^{m+1}\\
\varphi_2^{m+1}\\
\vdots\\
\varphi_{N-1}^{m+1}\\
\varphi_{N}^{m+1}
\;
\end{array}
\right\|
=
\left\|
\begin{array}
[c]{ll}%
A_1\\
A_2\\
\vdots\\
A_{N-1}\\
A_{N}
\;
\end{array}
\right\|,
\label{TriMatr}
\end{equation}
where
\begin{eqnarray}
A_n=b_{n,1} f^{m}_{n-1}+b_{n,2} \widetilde{f}^{m}_{n}+b_{n,3} f^{m}_{n+1}+c_{n,1} \varphi^{m-1}_{n-1}+c_{n,2} \varphi^{m-1}_{n}+c_{n,3} \varphi^{m-1}_{n+1}&
\\
for\hspace{0.5cm}n=1,...,N-1,\hspace{0.5cm}m=1,2,...M,\nonumber\\ \nonumber\\
A_1=b_{1,1}I_i^m+b_{1,2}\widetilde{f}^{m}_{1}+b_{1,3}f^{m}_{2}+
c_{1,2}\varphi^{m-1}_{1}+c_{1,3}\varphi^{m-1}_{2}+C_{1}^{-}I_{\text{bias}} &
\nonumber\\ for\hspace{0.5cm}n=0,\hspace{1.25cm}m=1,2,...M,\nonumber\\ \nonumber\\
A_{N}=b_{N,1}f^{m}_{N-1}+b_{N,2}\widetilde{f}^{m}_{N}+b_{n,3}I^{m}_\ell+
c_{N,1}\varphi^{m-1}_{N-1}+c_{N,2}\varphi^{m-1}_{N}+I_{\text{bias}}.&\nonumber\\
for\hspace{0.5cm}n=N,\hspace{0.50cm}m=1,2,...M. \nonumber\\
\nonumber
\end{eqnarray}
\end{widetext}
Equations~\eqref{TriMatr} are solved through a tridiagonal algorithm, which is a simplified form of Gaussian elimination. 
The solutions correspond to the Josephson phases $\varphi_n^{m+1}$, while the currents flowing through the input resistance $I_i$ and the load resistance $I_\ell$ are given by
\begin{widetext}
\begin{eqnarray}
I_i^{m+1}&=&I_i^{m-1}+\left [ \varphi^{m+1}_1\alpha_{1}^{+}-I_i^m+f_1^m+\varphi^{m-1}_1\alpha_{1}^{-}+\left (C_i\dot{V}_i^m-I_{\text{bias}}\right ) \right ]\left ( 2k\omega_i \right )\\
 I_\ell^{m+1}&=&I_\ell^{m-1}+\Bigg [f_N^m-\left (1+\frac{C_N}{C_\ell}\right )I^m_\ell -\varphi^{m+1}_N\alpha_{N}^{+}+\varphi^{m-1}_N\alpha_{N}^{-}-I_{\text{bias}}\Bigg ]\frac{2k}{C_N R_\ell}.
\end{eqnarray}
\end{widetext}

The initial conditions chosen in the simulation are $\varphi_n^{m}=0$ and $\varphi_n^{m-1}=0$ for $n=1,2...N$.

The coefficients $a_{i,j}$, $b_{i,j}$, and $c_{i,j}$, with $i,j\in[1,N]$, are arranged in square tridiagonal matrices as
\begin{equation}
\begin{bmatrix}
x_{1,2} & x_{1,3} & 0 & \ldots & 0 \\
x_{2,1} & x_{2,2} & x_{2,3} & \ldots & 0 \\
\vdots & \ddots & \ddots & \ddots & \vdots \\
0 & \ldots & x_{N-1,1} & x_{N-1,2} & x_{N-1,3} \\
0 & \ldots & 0 & x_{N,1} & x_{N,2}
\end{bmatrix},
\end{equation}
where $x \equiv \{a, b, c\}$, which elements are listed in Eqs.~\eqref{Eq20}-\eqref{bcl}-\eqref{bcr}, and depends on the vectors
\begin{equation}
\begin{aligned}
&\widetilde{C}_{J,n}=C_{J,n}\frac{\hbar}{2e} \qquad \frac{1}{\widetilde{R}_{J,n}}=\frac{1}{R_{J,n}}\frac{\hbar}{2e} \\
&\frac{1}{\widetilde{L}_{n}}=\frac{1}{L_{n}}\frac{\hbar}{2e} \qquad \widetilde{C}_{n}=C_{n}\frac{\hbar}{2e} \qquad C_{n}^{-}=\frac{C_{n}}{C_{n-1}} \\
&\widetilde{C}_{\text{``n''}}=\frac{\hbar}{2e}\left(C_{J,n} + \frac{C_{n-1}C_{n}}{C_{n-1}+C_{n}}\right) \\
&\alpha_n^{\pm}=\frac{\widetilde{C}_{J,n}}{k^2}\pm\frac{1}{2\widetilde{R}_{J,n}k} \qquad\widetilde{\alpha}_n^{\pm}=\frac{\widetilde{C}_{\text{``n''}}}{k^2}\pm\frac{1}{2\widetilde{R}_{J,n}k},
\end{aligned}
\end{equation}
with $n\in[1,N]$ and $C_0\equiv C_i$.

\begin{comment}
\begin{widetext}
\begin{footnotesize}
\begin{equation}
\mathcal{A}\!=\!\left\|
\begin{array}{@{} *{5}{l} @{}}
a_{1,2} & a_{1,3} & 0 & \ldots & 0 \\
a_{2,1} & a_{2,2} & a_{2,3} & \ldots & 0 \\
\vdots & \ddots & \ddots & \ddots & \vdots \\
0 & \ldots & a_{N-1,1} & a_{N-1,2} & a_{N-1,3} \\
0 & \ldots & 0 & a_{N,1} & a_{N,2}
\end{array}
\right\|\!,\;\;
\mathcal{B}\!=\!\left\|
\begin{array}{@{} *{5}{l} @{}}
b_{1,2} & b_{1,3} & 0 & \ldots & 0 \\
b_{2,1} & b_{2,2} & b_{2,3} & \ldots & 0 \\
\vdots & \ddots & \ddots & \ddots & \vdots \\
0 & \ldots & b_{N-1,1} & b_{N-1,2} & b_{N-1,3} \\
0 & \ldots & 0 & b_{N,1} & b_{N,2}
\end{array}
\right\|\!,\;\;
\mathcal{C}\!=\!\left\|
\begin{array}{@{} *{5}{l} @{}}
c_{1,2} & c_{1,3} & 0 & \ldots & 0 \\
c_{2,1} & c_{2,2} & c_{2,3} & \ldots & 0 \\
\vdots & \ddots & \ddots & \ddots & \vdots \\
0 & \ldots & c_{N-1,1} & c_{N-1,2} & c_{N-1,3} \\
0 & \ldots & 0 & c_{N,1} & c_{N,2}
\end{array}
\right\|\!,\!\!\!
\end{equation}
\end{footnotesize}
\end{widetext}
\end{comment}

%\bibliographystyle{apsrev4-1}
%\bibliography{biblio.bib}


\end{document}